\documentclass[12pt,a4paper]{article}
\usepackage[utf8x]{inputenc}
\usepackage{jheppub}
\usepackage{rotating}
\usepackage{bm}        
\usepackage{amssymb}   
\usepackage{enumitem}

\def\be{\begin{equation}}
\def\ee{\end{equation}}
\def\beal{\begin{equation}\begin{aligned}}
\def\eeal{\end{aligned}\end{equation}}
\def\nn{\nonumber}

\def\bra#1{\langle #1|}
\def\ket#1{|#1 \rangle}
\def\braket#1{\langle #1 \rangle}
\def\la{\lambda}
\def\lb{\tilde{\lambda}}

\def\ie{i.e. }
\def\eg{e.g. }

\def\eqn#1{eq.~\eqref{#1}}
\def\eqns#1#2{eqs.~\eqref{#1} and~\eqref{#2}}

\def\fig#1{figure~{\ref{#1}}}

\def\sec#1{section~{\ref{#1}}}
\def\secs#1#2{sections~{\ref{#1}} and~{\ref{#2}}}

\def\app#1{appendix~{\ref{#1}}}

\def\rcites#1{refs.~\cite{#1}}

\newcommand{\1}{\bm{1}}
\newcommand{\2}{\bm{2}}
\newcommand{\rma}{{\rm a}}

\newcommand{\rin}{\text{in}}

\subheader{\normalsize
	\vspace*{-3.7em}
	\begin{flushright}
		IRMP-CP3-23-35\\
	\end{flushright}
}

\title{Gravitational partial-wave absorption from scattering amplitudes}
\author[1]{Rafael Aoude}
\author[2]{and Alexander Ochirov}

\affiliation[1]{Centre for Cosmology, Particle Physics and Phenomenology (CP3), \\
Universit\'e catholique de Louvain, 1348 Louvain-la-Neuve, Belgium}
\affiliation[2]{London Institute for Mathematical Sciences, Royal Institution, 21 Albemarle Street, \\
London W1S 4BS, UK}

\emailAdd{rafael.aoude@uclouvain.be}
\emailAdd{ao@lims.ac.uk}

\abstract{We study gravitational absorption effects using effective on-shell scattering amplitudes. We develop an in-in probability-based framework involving plane- and partial-wave coherent states for the incoming wave to describe the interaction of the wave with a black hole or another compact object. We connect this framework to a simplified single-quantum analysis. The basic ingredients are mass-changing three-point amplitudes, which model the leading absorption effects and a spectral-density function of the black hole. As an application, we consider a non-spinning black hole that may start spinning as a consequence of the dynamics. The corresponding amplitudes are found to correspond to covariant spin-weighted spherical harmonics, the properties of which we formulate and make use of. We perform a matching calculation to general-relativity results at the cross-section level and derive the effective absorptive three-point couplings. They are found to behave as ${\cal O}(G_\text{Newton}^{s+1})$, where $s$ is the spin of the outgoing massive state.
}

\begin{document}
\maketitle
\addtocontents{toc}{\protect\setcounter{tocdepth}{2}}

\section{Introduction}
\label{sec:Intro}

Gravitational-wave observations~\cite{LIGOScientific:2016aoc} have been stimulating
the search of new computational methods for general relativity (GR).
In addition to classical approaches to the gravitational two-body problem, which have seen constant improvement~\cite{Mino:1996nk,Quinn:1996am,Jaranowski:1997ky,Damour:1999cr,Buonanno:1998gg,Buonanno:2000ef,Goldberger:2004jt,Porto:2016pyg,Damour:2016gwp,Damour:2017zjx,Levi:2018nxp},
new results have been obtained via scattering amplitudes, which encode
the relevant, quantum or classical, physics in a gauge-invariant way. See~\cite{Travaglini:2022uwo,Bern:2022wqg,Bjerrum-Bohr:2022blt,Kosower:2022yvp} for recent reviews.

The conservative scattering of non-spinning compact bodies has been calculated
up to fourth post-Minkowskian (PM) order using amplitude-~\cite{Bern:2021yeh,Damgaard:2023ttc} and worldline-based methods~\cite{Dlapa:2021npj,Dlapa:2021vgp,Dlapa:2022lmu,Jakobsen:2023ndj}.
For the spinning case, the conservative scattering has been evaluated at second PM order and all-order in the angular momenta~\cite{Aoude:2022trd,Aoude:2022thd,Aoude:2023vdk} with the help of Heavy-Particle Effective Theory~\cite{Damgaard:2019lfh,Aoude:2020onz}.
Higher PM orders have also been obtained,
though limited to lower spin orders~\cite{Mogull:2020sak,Jakobsen:2021smu,Jakobsen:2021lvp,FebresCordero:2022jts,Jakobsen:2023ndj}.
Progress on the spinning front has resulted in different and complementary on-shell approaches~\cite{Guevara:2018wpp,Chung:2018kqs,Guevara:2019fsj,Chung:2019duq,Aoude:2020onz,Bern:2020buy,Liu:2021zxr,Bautista:2021wfy,Bautista:2022wjf,Jakobsen:2022fcj,Bern:2022kto,Alessio:2023kgf}.
For the interesting case of black-hole (BH) dynamics,
many of these works rely on the matching~\cite{Guevara:2018wpp,Chung:2018kqs} of three-point amplitudes
to the Kerr multipole expansion~\cite{Vines:2017hyw}.
An all-spin understanding of the relevant
four-point Compton scattering amplitude, however, is still lacking,
despite recent progress in the description of massive higher-spin particles~\cite{Chiodaroli:2021eug,Ochirov:2022nqz,Cangemi:2022bew},
matching to the Teukolsky-equation solutions~\cite{Bautista:2021wfy,Bautista:2022wjf}
through sixth order in spin,
and the availability of the conservative tree-level Compton
with arbitrary coefficients~\cite{Haddad:2023ylx}.
The quantum-field-theoretic (QFT) program of gravitational dynamics
has also seen impressive advances in 
methods for obtaining classical observables from amplitudes,
such as the Kosower-Maybee-O'Connell (KMOC) formalism \cite{Kosower:2018adc,Maybee:2019jus,delaCruz:2020bbn,Cristofoli:2021vyo,Aoude:2021oqj,Cristofoli:2021jas}, heavy-particle expansion~\cite{Damgaard:2019lfh,Aoude:2020onz,Haddad:2020tvs,Brandhuber:2021kpo,Brandhuber:2021eyq,Brandhuber:2021bsf,Brandhuber:2023hhy},
eikonal ideas~\cite{Bern:2020gjj,Parra-Martinez:2020dzs,DiVecchia:2020ymx,DiVecchia:2021bdo,Adamo:2021rfq,Alessio:2022kwv,DiVecchia:2022nna},
worldline QFT~\cite{Mogull:2020sak,Jakobsen:2021smu,Jakobsen:2021lvp}, boundary-to-bound map~\cite{Kalin:2019rwq,Kalin:2019inp,Cho:2021arx,Gonzo:2023goe}, and strong-field amplitudes~\cite{Adamo:2022rmp,Adamo:2022qci,Adamo:2023cfp}.

Despite the successes in the conservative section, the progress in non-conservative effects has been slower, since those effects are naturally smaller.
In particular, the absorption of mass and angular momentum is tiny, especially for non-spinning bodies, and is unlikely to be observed by ground-based detects, as shown in~\cite{Alvi:2001mx} for 5-to-50 solar masses black holes.
However, for space-based detectors, the fraction of the radiated energy that is absorbed by the BHs will be around 5\%~\cite{Poisson:2004cw}.
This becomes especially important for rapidly rotating BHs, as shown in~\cite{Hughes:2001jr}.
The change of mass and spin of a BH naturally leads to a change in the horizon by the second law of BH thermodynamics~\cite{Hawking:1971vc}.
Such effects are already included in a few of the effective-one-body waveform templates~\cite{Nagar:2011aa,Bernuzzi:2012ku,Taracchini:2013wfa} and will be needed for a future precision program.

In this paper, we initiate the study of absorption effects using modern on-shell methods for scattering amplitudes.
In particular, we use mass-changing three-point amplitudes to describe leading absorption effects from a simplified single-quantum approach.
We thus construct an in-in on-shell probability-based formalism for a partial wave impinging on a BH.
Using this covariant effective-field-theory (EFT) description,
we can match the microscopic cross-section calculation from GR literature
and obtain the values of the relevant effective coupling coefficients.
As a concrete application, we focus on absorption by a non-spinning BH,
while leaving the more phenomenologically relevant spinning case
for future work.

Absorption effects have been considered before in the literature,
starting with Starobinsky, Churilov~\cite{Starobinsky:1973aij,Starobinskil:1974nkd} and Page~\cite{Page:1976df,Page:1976ki}
and with later higher-order corrections in~\cite{Zhang:1985qz} and relatively recently in~\cite{Dolan:2008kf} by using traditional GR methods.
The scattering and absorption cross-sections are obtained
using a partial-wave expansion (in spin-weighted spherical harmonics)
of the scattering phases and transmission factors.
These factors are obtained by solving the Teukolsky equation, which describes perturbation around Kerr BHs.
Absorption of mass and angular momentum by a BH was also computed in great detail~\cite{Poisson:1993vp,Poisson:1994yf,Alvi:2001mx,Poisson:2004cw}
in post-Newtonian theory.

From the worldline perspective, the study of absorption is more recent.
It has been considered in~\cite{Goldberger:2005cd} for scalar BHs,
with subsequent inclusion of spin effects~\cite{Porto:2007qi,Goldberger:2020fot}.
Furthermore, absorption has been combined with spontaneous emission to understand superradiance effects in~\cite{Endlich:2016jgc}.
The authors of \cite{Goldberger:2005cd,Porto:2007qi,Endlich:2016jgc,Goldberger:2020fot,Saketh:2022xjb} put EFT operators on a classical worldline
to model the intricate behavior of a compact object.
In particular, higher-derivative operators were included in~\cite{Saketh:2022xjb} for the spinning case, which starts at 1.5 PN order, tackling the discrepancy in the literature of the horizon fluxes in the test-body limit.
We propose to go further and consider the object itself as a quantum particle,
but amenable to an appropriately defined classical limit.
This lets us profit not just from QFT techniques,
which have been available on the worldline,
but also from the on-shell approach to scattering amplitudes.

Purely mass-changing absorption effects from on-shell scattering amplitudes were never studied to the best of our knowledge,\footnote{In \cite{Bautista:2021wfy,Bautista:2022wjf} the authors have introduced contact terms non-analytical in spin for the Compton amplitude to match the solutions of the Teulkosky equation. These terms are then suggested to model absorption effects, despite the masses of the initial and the final particles being equal. Here what we call absorption effects are strictly inelastic changing-mass interactions.
}
although similar amplitudes have appeared in the context of quantum emission~\cite{Kim:2020dif,Kim:2020cvf}.
The basic building blocks for modeling absorption effects
are three-point amplitudes of two different massive particles and a graviton,
in which the initial state absorbs the graviton,
changing its mass and spin. These amplitudes induce the introduction of a spectral-density function for the black holes, which goes beyond the simplest point particle approach.
Even before matching, the EFT cross-section reproduces
known facts about Schwarzchild BHs:
\textit{(i)} the cross-section does not depend on the magnetic quantum number $m$, and
\textit{(ii)} there is no absorption in the static limit $\sigma_\text{abs} (\omega_\text{cl} \to 0) = 0$. 

Properly modeling the interaction of a BH with a classical wave
from amplitudes requires the use of massless coherent states.
For that, we describe a covariant probability-based formalism for spherical coherent states,
so as to substantiate the single-quantum leading-order calculation, and to explain how one could improve the absorption description to higher orders and combine it with conservative effects.

This paper is organized as follows.
In \sec{sec:Mechanics} we give construct a scattering-matrix element for a compact object absorbing a partial/spherical wave in terms of mass-changing and spin-changing amplitudes, the form of which is specified in \sec{sec:3pts}.
In \sec{sec:Absorption3pts} we combine these ingredients into an absorptive cross-section
with the help of the BH spectral density function, which enters as an additional effective coupling factor.
In~\sec{sec:Matching} we match to the microscopic cross-section from GR to make sense of the effective couplings.
Finally, in \sec{sec:CoherentXsec}, we connect the single-quantum cross-section description to the framework involving massless spherical coherent states. In this section, we also introduce a diagrammatic expansion of the $T$-matrix, which allows for perturbations of the BH-wave interaction that can be matched to higher orders of the cross-section.
We conclude in \sec{sec:Outro}.
Though we assume familiarity with the spinor-helicity formalism \cite{Arkani-Hamed:2017jhn},
we briefly explain it and its connection to covariant spin-weighted spherical harmonics in \app{app:SphHarmCov}.

\section{Partial-wave absorption matrix element}
\label{sec:Mechanics}

In this section we describe our setup for obtaining
classical absorption effects from the quantum on-shell scattering amplitudes.
Such a setup of course relies on EFT ideas, such as treating black holes as point particles.
These concepts have been heavily used recently to provide predictions for conservative dynamics and dissipation effects.

As in most of EFTs, the knowledge of the coefficients that parametrize the theory is either provided by experimental data or by performing a matching calculation to the underlying theory.
In our case, the underlying theory is Einstein's GR, or more practically, the solution to the Teukolsky equation
\cite{Teukolsky:1973ha,Press:1973zz,Teukolsky:1974yv}.
Given these two sides of the EFT matching, we will sometimes be referring to the EFT side of the calculation as macroscopic, and to the solution to Teukolsky equation as microscopic.
On the EFT side, we will model absorption effects using mass-changing amplitudes.

\begin{figure}[t]
	\centering
	\vspace{-15pt}
	\includegraphics[width=9cm]{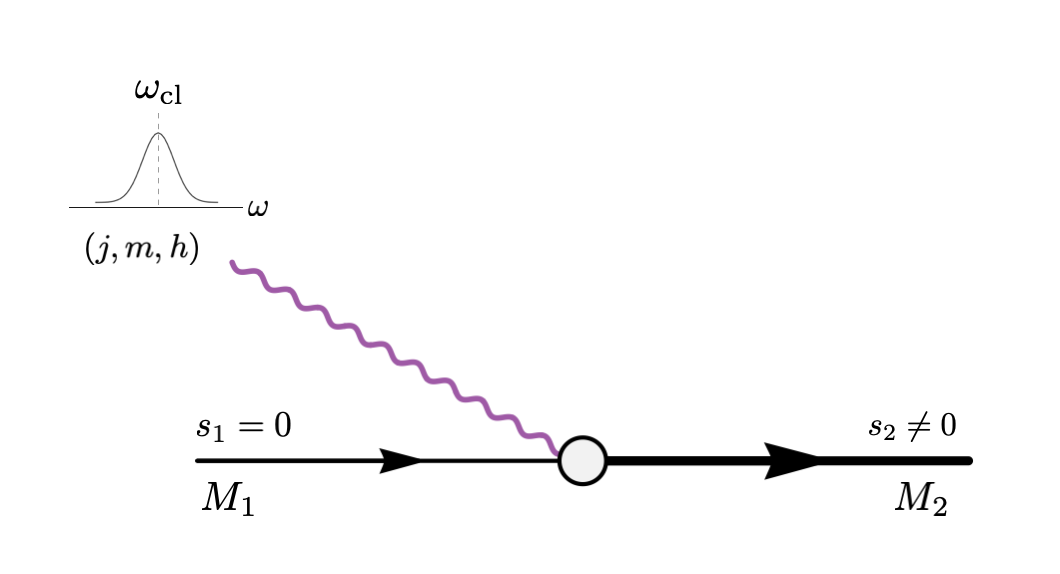}
	\vspace{-15pt}
	\caption{Wave impinging on a scalar black hole}
	\label{fig:absorption}
\end{figure}

We focus on the simplest relevant process depicted in \fig{fig:absorption}:
a graviton spherical state
impinging on a massive particle of mass~$M_1$ (for simplicity taken spinless),
which absorbs the graviton and changes its mass to $M_2$ and spin to $s_2$.
It is natural to think of the corresponding scattering amplitude
in terms of plane-wave states as described in \sec{sec:3pts}.
However, GR methods give us results~\cite{Starobinsky:1973aij,Starobinskil:1974nkd,Page:1976df,Page:1976ki,Dolan:2008kf,Chia:2020yla,Charalambous:2021mea,Ivanov:2022qqt}
for spherical waves with fixed angular-momentum quantum numbers.
Therefore, we start by translating between these two pictures
--- with a focus on single-graviton states.
In \sec{sec:CoherentXsec} we will come back to justifying this setup further
using classical coherent states,
which are more appropriate for modeling classical waves.

\subsection{Spherical helicity amplitude}
\label{sec:SphHelicity}

By definition (see \eg \cite{Berestetskii:1982qgu}), spherical helicity states partially diagonalize the total spin operator~$\bm{J}$.
They are eigenstates of $\bm{J}^2$, $J_z$ and helicity $(\bm{J}\cdot \bm{P})/\bm{P}^2$, as well as the Hamiltonian~$P^0$.
These states are labeled by energy~$\omega$,
angular-momentum quantum numbers $j$, $m= -j, \dots, j$
and helicity $h=\pm 2$ (graviton) or $\pm 1$ (photon):\footnote{Here and below,
   the hat notation~\cite{Kosower:2018adc} means
   $\hat{d}^n p :=  d^np /(2\pi)^n$ and
   $\hat{\delta}^n(...) := (2\pi)^n \delta^n(...)$.
   For the spherical helicity states, we also assume that masslessness:
   $P^2 \ket{\omega,j,m,h} = 0$.
}
\be
\ket{\omega,j,m,h} = a_{j,m,h}^\dagger(\omega) \ket{0},	\qquad
\braket{\omega',j',m',h'|\omega,j,m,h} = \hat{\delta}(\omega'-\omega)
\delta^j_{j'} \delta^m_{m'} \delta^h_{h'} .
\label{eq:SphHelStates}
\ee
This is in contrast to the more familiar plane-wave states~$\ket{k,h}$,
which diagonalize the four-momentum $P^\mu$
in addition to the helicity $(\bm{J}\cdot \bm{P})/\bm{P}^2$:
\be
\ket{k,h} := a_h^\dagger(k) \ket{0}, \qquad
\braket{k',h'| k,h} = 2|\bm{k}| \hat{\delta}^3(\bm{k}'-\bm{k})
\delta^h_{h'} .
\label{eq:PlainWaveStates}
\ee
The two bases of one-particle states may be related by~\cite{Endlich:2016jgc}
\be
\braket{k,h'|\omega,j,m,h} = \frac{4\pi}{\sqrt{2\omega}}
\delta^h_{h'} \hat{\delta}(|\bm{k}|-\omega)\,{}_{-h}Y_{jm}(\hat{\bm{k}}) ,
\ee
where the spin-weighted spherical harmonics ${}_{-h}Y_{jm}(\hat{\bm{k}})$
depend on the momentum direction $\hat{\bm{k}} := \bm{k}/|\bm{k}|$
and constitute a generalization \cite{Newman:1966ub,Goldberg:1966uu}
of the usual (scalar) spherical harmonics.
The corresponding completeness relations imply
that the one-particle spinning spherical state can be written as
\be\!\!\!
\ket{\omega,j,m,h}\;\!\!
=\;\!\!\frac{4\pi}{\sqrt{2\omega}}\!\int_k\!\hat{\delta}(k^0\!-\omega)\,
{}_{-h}Y_{j,m}(\hat{\bm{k}}) \ket{k,h}\;\!\!
=\;\!\!\sqrt{2\omega}\;\!\!\!\int\;\!\!\!\frac{d\Omega_{\hat{\bm{k}}}}{4\pi}
{}_{-h}Y_{j,m}(\hat{\bm{k}}) \ket{k,h} \big|_{|\bm{k}|=\omega} ,\!
\label{eq:SphHel2PlaneWave}
\ee
where $d\Omega_{\hat{\bm{k}}}$ denotes the spherical-angle integration measure
over the directions of~$\bm{k}$.
Here and below,
we use a shorthand for the on-shell integration measure (for $M_k=0$)
\be
\int_p := \int\!\!\frac{d^4 p}{(2\pi)^3} \Theta(p^0) \delta(p^2-M_p^2)
=:\!\int\!\!\frac{d^4 p}{(2\pi)^3} \delta^+(p^2-M_p^2) .
\label{eq:OnshellMeasure}
\ee

In order to write the scattering matrix element for a spherical helicity state,
we need to be careful with the massive particle at the origin,
which, strictly speaking, cannot be a plane-wave state either.
So instead we use a wavepacket
\be
\ket{\psi} := \int_{p_1}\!\!\psi_\xi(p_1) \ket{p_1} : \qquad
\begin{aligned}
 & \braket{\psi|\psi} = 1 , \qquad
	\braket{\psi|P^\mu|\psi} = p_{1,\text{cl}}^\mu := (M_1,\bm{0}) , \\
 & \braket{\psi|P^\mu P^\nu|\psi} = \braket{\psi|P^\mu|\psi} \braket{\psi|P^\nu|\psi}
	+ {\cal O}(\xi) .
\end{aligned}
\label{eq:Wavepacket}
\ee
For concreteness, we may think of
$\psi_\xi(p_1) \propto \exp\big({-}\tfrac{p_1^0}{\xi M_1}\big)$.
The dimensionless parameter $\xi = \ell_\text{C}^2/\ell_\text{WP}^2$
encodes the ratio of the Compton wavelength and the position-space spread of the wavepacket
\cite{Kosower:2018adc}.
We will be focusing on the scale hierarchy
$\ell_\text{C} \ll \ell_\text{WP} \ll 2\pi \hbar c/\omega$
relevant for classical scattering of a wave with frequency $\omega/\hbar$.

We are now ready to express the $S$-matrix element
for a spherical helicity state in terms of the conventional plane-wave scattering amplitude:
\be
\braket{X|S|\psi;\omega,j,m,h} = \frac{4\pi i}{\sqrt{2\omega}}\!
\int_{p_1}\!\!\psi_\xi(p_1)\!\int_k\!\hat{\delta}(k^0\!-\omega)
\hat{\delta}^{4}(p_1+k-p_X)\,{}_{-h}Y_{j,m}(\hat{\bm{k}})
{\cal A}(X|p_1;k,h) .
\label{eq:SphHelAmplitude}
\ee
As usual, we have ignored the no-scattering term in $S=1+iT$.
For the amplitude arguments,
we choose to mimic the structure of the matrix elements
and write the outgoing particles first separated
from the incoming particles by a vertical line.

Unfortunately, the matrix element~\eqref{eq:SphHelAmplitude} by itself
is too singular to handle unambiguously, which is due to the infinite norm
$\braket{\omega,j,m,h|\omega,j,m,h} = \hat{\delta}(0)$
of the massless spherical state~\eqref{eq:SphHel2PlaneWave}.
So we also smear its energy with a wavefunction:
\be
\ket{\gamma} := \int_0^\infty\!\!\hat{d}\omega\:\!
\gamma_\zeta(\omega) \ket{\omega,j,m,h} : \qquad
\begin{aligned}
	& \braket{\gamma'|\gamma}
	= \delta^j_{j'} \delta^m_{m'} \delta^h_{h'} , \qquad
	\braket{\gamma|P^0|\gamma} = \omega_\text{cl} , \\
	& \braket{\gamma|P^0 P^0|\gamma} = \braket{\gamma|P^0|\gamma} \braket{\gamma|P^0|\gamma}
	+ {\cal O}(\zeta) .
\end{aligned}
\label{eq:WavepacketSph}
\ee
The corresponding scattering-matrix element is finally
\be
\braket{X|S|\psi;\gamma}
= 4\pi i\!\int_{p_1}\!\!\psi_\xi(p_1)\!\int_k
\frac{\gamma_\zeta(k^0)}{\sqrt{2k^0}}
\hat{\delta}^{4}(p_1+k-p_X)\,{}_{-h}Y_{j,m}(\hat{\bm{k}})
{\cal A}(X|p_1;k,h) .
\label{eq:SphHelAmplitude2}
\ee

\subsection{Covariant spherical states}
\label{sec:LOabsXsectionCov}

Before we proceed to the absorption cross-section,
it is rewarding to covariantize our spherical-helicity state setup.
By covariantization we mean allowing for an arbitrary time direction $u^\mu$,
with $u^2=1$, as well a spacelike spin quantization axis $n^\mu$,
with $n^2=-1$ and  $n \cdot u = 0$.
(In \sec{sec:SphHelicity}, these were set to $(1,\bm{0})$ and $(0,0,0,1)$, respectively.)
The corresponding angular-momentum operator is
\be
J^\mu(u) := \frac{1}{2} \epsilon^{\mu\nu\rho\sigma} J_{\nu\rho} u_\rho
\qquad \Rightarrow \qquad
[J^\mu(u), J^\nu(u)] = i\epsilon^{\mu\nu\rho\sigma} u_\rho J_\sigma(u) ,
\label{eq:AngMomCov}
\ee
which is not to be confused with the Pauli-Lubanski spin vector $W^\mu := \epsilon^{\mu\nu\rho\sigma} J_{\nu\rho} P_\rho/2$.
A spherical helicity state $\ket{\omega,j,m,h}$ is then an eigenstate
of ``energy'' $E(u):= u \cdot P$ and angular-momentum combinations
$-J(u)^2$, $n \cdot J(u)$ and $J(u) \cdot P = W \cdot P$.
Similarly to \eqn{eq:SphHel2PlaneWave}, we choose to construct them
directly from the plane-wave states:
\be
\ket{\omega,j,m,h}_{u,n}
:= \frac{4\pi}{\sqrt{2\omega}}\!\int\!\hat{d}^4k\,
\hat{\delta}^+(k^2) \hat{\delta}(k \cdot u -\omega)\,
{}_{-h}Y_{j,m}(k;u,n) \ket{k,h} .
\label{eq:SphHel2PlaneWaveCov}
\ee

The crucial new ingredient here is the covariant spin-weighted spherical harmonic.
We define these functions in terms of spinor products as follows:
\beal
{}_h\tilde{Y}_{j,m}(k;u,n) := &\;
\frac{1}{\bra{k}u|k]^j}\,
\Big[ [u_a k]^{\odot (j+h)}
\odot \braket{k u_a}^{\odot (j-h)}
\Big]_{\{a\} = ({\scriptsize\underbrace{1 \dots 1}_{j-m}
		\underbrace{2 \dots 2}_{j+m}})} .
\label{SphHarmSpinors4d}
\eeal
We have hereby followed \cite{Guevara:2021yud,Bautista:2022wjf}
in using the massive spinor-helicity formalism~\cite{Arkani-Hamed:2017jhn}
(see \app{app:SphHarmCov} for a brief review)
to covariantize the spinorial construction
dating back to Newman and Penrose~\cite{Newman:1966ub}.
The exposed indices~$\{a\}=\{a_1,\dots,a_{2j}\}$ correspond to the little group ${\rm SU}(2)$ of $u^\mu$
and are understood to be fully symmetrized, as indicated by the symmetric tensor-product symbol~$\odot$.
We adopt the spinor conjugation conventions
$\big( \ket{p^a}_\alpha \big)^*\!=[p_a|_{\dot{\alpha}}$,
$\big( [p^a|_{\dot{\alpha}} \big)^*\!=-\ket{p_a}_\alpha$ for $p^0>0$,
which immediately imply
\be
{}_h\tilde{Y}_{j,m}^*(k;u,n) = (-1)^{2j+m-h} {}_{-h}\tilde{Y}_{j,-m}(k;u,n) .
\label{SphHarmSpinors4dConj}
\ee
The properly normalized functions seen in \eqn{eq:SphHel2PlaneWaveCov}
are written without the tildes:
\be
{}_hY_{j,m}(k;u,n) := (-1)^m (2j)!
\textstyle\sqrt{\frac{2j+1}{4\pi (j+m)!(j-m)! (j+h)!(j-h)!}}\;
{}_h\tilde{Y}_{j,m}(k;u,n) ,
\label{SphHarmSpinors4dNorm}
\ee
with the orthonormality statement being
\be
\frac{2}{\omega}\!\int\!d^4 k\;\!\delta^+(k^2) \delta(k \cdot u - \omega)\,
{}_hY^*_{j',m'}(k;u,n)\;{}_hY_{j,m}(k;u,n)
= \delta_j^{j'} \delta_m^{m'} .
\label{SphHarmCovNorm}
\ee
For the proof and a more detailed exposition of the harmonics~\eqref{SphHarmSpinors4d},
see \app{app:SphHarmCov}.

Here let us point out the new important features of these harmonics.
First of all, the harmonics are by definition~\eqref{SphHarmSpinors4d}
insensitive to the overall scale of both~$k^\mu$ and~$u^\mu$.
Moreover, they are now clearly formulated in a convention-independent way ---
in the sense that it is covariant with respect to the two little groups:
\begin{itemize}[leftmargin=21pt,parsep=0pt]
	\item
	the massless little-group ${\rm U}(1)$ of $k^\mu$ may be used to change
	the phases of all spherical harmonics in a local but mutually consistent way.
	Namely, transforming
	$\ket{k} \to e^{-i\phi(k)/2} \ket{k}$, $|k] \to e^{i\phi(k)/2} |k]$
	implies phase adjustments of the form
	${}_hY_{j,m}(k;u,n) \to e^{ih\phi(k)} {}_hY_{j,m}(k;u,n)$,
	which connect between various possible definitions
	of spin-weighted spherical harmonics,
	\eg via quaternions~\cite{Boyle:2016tjj}.
	
	\item
	the massive little group ${\rm SU}(2)$ of $u^\mu$ may be used to change
	the physical meaning of the magnetic quantum number~$m$.
	For instance,
	the explicit spinor parametrizations~\eqref{MassiveSpinorSolutionAngles}
	and~\eqref{MassiveSpinorSolutionRest}
	correspond to the $m$-quantization along~$\bm{u} \neq 0$
	and the conventional $z$-axis for $\bm{u}=0$, respectively.
	However, we may just as well apply transformations
	$\ket{u^a} \to U^a{}_b(u) \ket{u^b}$,
	$|u^a] \to U^a{}_b(u) |u^b]$
	to the massive spinors,
	and this will rotate the spin quantization axis
	\be
	n^\mu := \frac{1}{2}
	(\bra{u_2}\sigma^\mu|u^2] + [u_2|\bar{\sigma}^\mu\ket{u^2})
	\qquad \Rightarrow \qquad
	n^2 = -1, \quad u \cdot n = 0 .
	\label{SpinQuantAxis}
	\ee
	Having this relation in mind,
	we henceforth compress our notation to ${}_hY_{j,m}(k;u)$.
\end{itemize}

In addition, we can specify general frame transformations
of the covariant spherical harmonics~\eqref{SphHarmSpinors4d}.
Indeed, it is shown in \app{app:Transformations}
that under the time-direction change
$u^\mu \to v^\mu = L^\mu{}_\nu(v\!\leftarrow\!u) u^\nu$
the massive spinors are boosted as follows:
\be\!\!\!
\ket{v^a} = \frac{\sqrt{\mu}}{\mu\!+\!1} 
|u\!+\!v|u^a] , \qquad
|v^a] = \frac{\sqrt{\mu}}{\mu\!+\!1} 
|u\!+\!v\ket{u^a} , \qquad
\mu := u \cdot v + \sqrt{(u \cdot v)^2\!-1} .\!
\label{eq:SpinorTransform}
\ee
Here we have assumed that the spin quantization axis
for the resulting time direction~$v^\mu$
is automatically $L^\mu{}_\nu(v\!\leftarrow\!u) n^\nu$,
\ie the boosted version of the original quantization axis $n^\mu$.
Of course, it can then be easily tweaked
by an additional little-group transformation of the resulting spinors
$\ket{v^a} \to U^a{}_b(v) \ket{v^b}$,
$|v^a] \to U^a{}_b(v) |v^b]$.

Given this covariant formulation of the spherical states,
we rewrite~\eqn{eq:SphHelAmplitude2} as
\begin{align} \label{eq:SphHelSMatrixCov}
\braket{X|S|\psi;\gamma} = 4\pi i\!
	\int_0^\infty\!\!\frac{\hat{d}\omega}{\sqrt{2\omega}}
	\gamma_\zeta(\omega) &\!
	\int\!\hat{d}^4p_1 \hat{\delta}^+(p_1^2\!-\!M_1^2)
	\psi_\xi(p_1) \\ \times\!
	\int\!\hat{d}^4k\:\!\hat{\delta}^+(k^2)
	\hat{\delta}(k \cdot u_1-\omega) &
	\hat{\delta}^{4}(p_1+k-p_X)\,{}_{-h}Y_{j,m}(k;u_1)
	{\cal A}(X|p_1;k,h) . \nn
\end{align}
This is the scattering matrix-element formula that we are going to use in the absorption cross-section calculation below.
For concreteness, we will employ the following Lorentz-covariant realization of the massive wavefunction $\psi_\xi$ is
\cite{AlHashimi:2009bb,Kosower:2018adc}
\be
\psi_\xi(p_1) = \frac{1}{M_1}\!
\left[\frac{8\pi^2}{\xi K_1(2/\xi)}\right]^{1/2}
\exp\!\bigg(\!{-}\frac{p_1\!\cdot u_1}{\xi M_1}\bigg) ,
\label{eq:MomentumWF}
\ee
where $K_1$ denotes the modified Bessel function of the second kind.

So far we have not specified the form of the scattering amplitude ${\cal A}(X|p_1;k,h)$.
In our EFT approach to non-conservative effects, it is natural to assume that the leading contribution to the absorption process comes from mass-changing three-point amplitudes, \ie $X$ is particle of mass $M_2$ and $s_2$.
We discuss such amplitudes next.

\section{Basic mass-changing amplitudes}
\label{sec:3pts}

In this section we construct the three-point amplitudes that serve as basic building blocks for modeling absorption effects.
As discussed above, these amplitudes involve a massless messenger particle and two particles with different masses.
They were first explored in~\cite{Conde:2016vxs} and covariantized in~\cite{Arkani-Hamed:2017jhn}.
Here we further reorganize the latter formulation,
while also using coherent-spin eigenvalues~\cite{Aoude:2021oqj},
which saturate the massive spin indices and thus act as a book-keeping device~\cite{Chiodaroli:2021eug}.

We continue to work in the massive spinor-helicity formalism
\cite{Arkani-Hamed:2017jhn}, which is briefly reviewed in \app{app:SphHarmCov}.
In this formalism, an amplitude~${\cal A}_{\{b\}}{}^{\{a\}}$ involving two massive particles carries
$2s_1$ symmetrized little-group ${\rm SU}(2)$ indices $a_1,\dots,a_{2s_1}$ for the incoming particle~$1$,
and $2s_2$ such indices $b_1,\dots,b_{2s_2}$ for the outgoing particle~$2$.
We choose to use the chiral basis of massive spinors (angle brackets)
for positive helicities
and the antichiral basis (square brackets) for negative helicities.
Since $\det\{ \ket{1^a}_\alpha \} = \det\{ |1^a]_{\dot{\alpha}} \} = M_1$
and $\det\{ \ket{2^b}_\beta \} = \det\{ \ket{2^b}_{\dot{\beta}} \} = M_2$,
we may proceed by stripping the spinors $\ket{1_a}$ and $\ket{2^b}$
for the positive messenger helicity
and $|1_a]$ and $|2^b]$ for negative helicity.
For instance, in the positive-helicity case we write
\be
{\cal A}_{\{b\}}{}^{\{a\}} (p_2,s_2|p_1,s_1;k,h)
=: A(k,h)^{\{\alpha\},\{\beta\}} 
 (\ket{1_a}_\alpha)^{\odot 2s_1}
 (\ket{2^b}_\beta)^{\odot 2s_2} ,
\ee
where $\odot$ denotes the symmetrized tensor product.
In addition to the ${\cal A}_{\{b\}}{}^{\{a\}}$
and $A^{\{\alpha\},\{\beta\}}$ objects,
a third perspective on the same amplitude is provided
by contracting massive spinors with auxiliary ${\rm SU}(2)$-spinor variables~\cite{Chiodaroli:2021eug},
\be
\ket{\bm{1}} := \ket{1_a} \alpha^a , \qquad \quad
\ket{\bar{\bm{2}}} := \ket{2^b} \tilde{\beta}_b ,
\label{eq:AuxSpinors}
\ee
which may serve as an extra handle on the spin quantization axis.\footnote{The auxiliary
   ${\rm SU}(2)$ spinors $\alpha^a$ and $\tilde{\beta}_b$ transform under the action of
   the little groups of $p_1$ and~$p_2$, respectively,
   and in this sense have an implicit dependence on their momenta.
   Moreover, in the coherent-spin framework they constitute eigenvalues of the spin-annihilation operators~\cite{Schwinger:1952dse,Aoude:2021oqj}.
}
We write the fully contracted amplitude in boldface
as a scalar in terms of the spinor-stripped one:
\be
\bm{{\cal A}}(p_2,s_2| p_1,s_1;k,h)
:= A(k,h)^{\{\alpha\},\{\beta\}} 
	(|\bm{1}\rangle^{\otimes 2s_1})_{\{\alpha\}}
	(|\bar{\bm{2}}\rangle^{\otimes 2s_2})_{\{\beta\}} ,
\ee
the advantage being that the index symmetrization is now entirely automatic.

\subsection{Classifying mass-changing amplitudes}
\label{sec:review3pt}

Going back to the stripped amplitude $A(k,h)_{\{\alpha\},\{\beta\}}$
with two sets of symmetrized ${\rm SL}(2,\mathbb{C})$ indices,
we may decompose it in the chiral-spinor basis of $\ket{k}$ and $|p_1|k]$.
Unlike the equal-mass case, these two spinors are linearly independent
(and there is no need for a helicity factor~\cite{Arkani-Hamed:2017jhn}),
because
\be
\bra{k}p_1|k] = 2p_1\!\cdot k = M_2^2 - M^2_1 \neq 0
\ee
due to momentum conservation $p_2=p_1+k$.
This equation also tells us about the possible dimensionful scales entering the three-point process from an EFT perspective,
which will have to be matched later.
We can either use the mass pair $(M_1,M_2)$ or $(M_1,2p_1\!\cdot k)$,
and in this work we are going to favor the latter.
For instance, we may use $M_1$ to absorb the mass dimension of the amplitude
and allow the EFT coefficients to depend on the dimensionless ratio
\be
w := \frac{2p_1\!\cdot k}{M_1^2} ,
\label{eq:DimensionlessRatio}
\ee
while expanding in terms of the dimensionless spinors
of helicity $-1/2$ and $1/2$:
\be
\la_\alpha := M_1^{-1/2} \ket{k}_\alpha , \qquad \quad
\mu_\alpha := M_1^{-3/2} p_{1,\alpha\dot{\beta}} |k]^{\dot{\beta}} \qquad \Rightarrow \qquad
\braket{\la \mu} = w .
\label{eq:SpinorBasisChiral}
\ee

Therefore, the most general stripped amplitude
involving two unequal masses and one massless positive-helicity particle
is schematically given by \cite{Conde:2016vxs,Arkani-Hamed:2017jhn}
\be
\label{eq:ThreePointStripped}
A(k,h)_{\{\alpha\},\{\beta\}} = M_1^{1-s_1-s_2}\!\sum_i 
   c^h_{(i),s_1,s_2}(w)
   [\la^{s_1+s_2-h} \mu^{s_1+s_2+h}]^{(i)}_{\{\alpha\},\{\beta\}} .
\ee
Here $i$ enumerates inequivalent tensor products
with the given spinorial index structure,
and their scalar coefficients $c^h_{i,s_1,s_2}(\omega)$ may depend on each spin and in the dimensionless ratio $w$.
Before we specify the relevant spinorial structures,
note that there are natural constraints that follow already from the form of \eqn{eq:ThreePointStripped}, such as
\be
s_1+s_2 \pm h \,\in\, \mathbb{Z}_{\geq 0} \qquad \Rightarrow \qquad
s_1+s_2 \geq |h| .
\label{eq:ObviousConstraint}
\ee
Moreover, there can clearly be no three-point amplitude
for one or three half-integer spins
--- in QFT this standard fact is usually derived
from the spin-statistics theorem.

We find it helpful to observe that the massless little-group dependence
may be completely factored out (in the tensor-product sense).
This leaves a polynomial in $\la$ and~$\mu$, which is independent of the massless helicity:
\beal
\![\la \mu \oplus \mu \la]_{\{\alpha\},\{\beta\}}^n
:= c_0 (\la^n)_{\alpha_1\dots\alpha_n} (\mu^n)_{\beta_1\dots\beta_n}
 + c_1 (\la^{n-1} \mu)_{\alpha_1\dots\alpha_n}
   (\mu^{n-1} \la)_{\beta_1\dots\beta_n} & \\ + \ldots
 + c_{n-1} (\la \mu^{n-1})_{\alpha_1\dots\alpha_n}
   (\mu \la^{n-1})_{\beta_1\dots\beta_n}
 + c_n (\mu^n)_{\alpha_1\dots\alpha_n} (\la^n)_{\beta_1\dots\beta_n} &,
\label{eq:TensorPolynomial3pt}
\eeal
where we have also omitted the $\otimes$ sign for brevity.
The exponent $n$ depends on the total-spin quantum numbers,
and in the amplitude each such term may have its own coefficient.
Without loss of generality, we consider $s_2\geq s_1$,
where we have two cases:
\begin{itemize}[leftmargin=21pt,parsep=0pt]
\item $s_2-s_1 \geq h$,
where we saturate the $s_1$ indices by the above polynomial,
while the remaining $s_2$ indices are accounted for by the tensor product,
which is unambiguously defined by the overall helicity weight.
The corresponding spinorial structures belong to the following
tensor power of a direct sum:
\be
[\la^{s_1+s_2-h} \mu^{s_1+s_2+h}]^{(i)}_{\{\alpha\},\{\beta\}} ~\in~
   [\la \mu \oplus \mu \la]_{\{\alpha\},\{\beta\}}^{2s_1}\,
   (\la^{s_2-s_1-h} \mu^{s_2-s_1+h})_{\{\beta\}} ;
\ee

\item $s_2-s_1 < h$,
where the polynomial~\eqref{eq:TensorPolynomial3pt} saturates
the number of $\la$'s, which is equal to $s_1+s_2-h$,
while the remaining $2h$ of $\mu$'s are unambiguously distributed among
the two massive particles. The spanning spinorial structure is thus
\be
[\la^{s_1+s_2-h} \mu^{s_1+s_2+h}]^{(i)}_{\{\alpha\},\{\beta\}} ~\in~
   [\la \mu \oplus \mu \la]_{\{\alpha\},\{\beta\}}^{s_1+s_2-h}\,
   (\mu^{s_1-s_2+h})_{\{\alpha\}} (\mu^{s_2-s_1+h})_{\{\beta\}} .
\ee
Note that in electromagnetism this case only occurs for $s_1=s_2$,
whereas in GR both $s_2 = s_1$ and $s_2 = s_1 + 1$ are possible.
\end{itemize}
In both cases, we have the polynomial with free coefficients
and the additional factor, which carries the massless helicity.
This factor completes the ${\rm SL}(2,\mathbb{C})$ indices
of either massive particle that are not accounted for by the polynomial,
and of course all $\alpha$'s and all $\beta$'s are implicitly symmetrized.

This analysis should be repeated for $s_1 \leq s_2$,
and the ${\rm SL}(2,\mathbb{C})$ can then be contracted
with the massive spinors (and auxiliary variables),
for which the Dirac equations
$|p_1\ket{\1} = M_1|\1]$ and $|p_2\ket{\bar{\2}} = M_2|\bar{\2}]$ hold.
In this way, we arrive at
\begin{subequations} \label{eq:ThreePointGeneral}
\be
\bm{{\cal A}}(p_2,s_2|p_1,s_1;k,h) =
\left\{
\begin{aligned}
 & \bm{F}^h_{s_1,s_2}\,
   \braket{\bar{\2}k}^{s_2-s_1-h} [\bar{\2}k]^{s_2-s_1+h}\!, \\
 & \bm{F}^h_{s_1,s_2}\,
   [\bar{\2}k]^{s_2-s_1+h} [k\1]^{s_1-s_2+h}\!, \\
 & \bm{F}^h_{s_1,s_2}\,
   \braket{\bar{\2}k}^{s_2-s_1-h} \braket{k\1}^{s_1-s_2-h}\!, \\
 & \bm{F}^h_{s_1,s_2}\, \braket{k \1}^{s_1-s_2-h} [k \1]^{s_1-s_2+h}\!,
\end{aligned}
\right. \quad
\begin{aligned}
   s_2-s_1 & \geq |h| ,\\
  |s_2-s_1| & < h , \\
  |s_2-s_1| & < -h , \\
   s_1-s_2 & \geq |h| .
\end{aligned}
\label{eq:ThreePointGeneralAmp}
\ee
where the factor $\bm{F}^h_{s_1,s_2}$
contains free coefficients and can now be written as
\be
\bm{F}^h_{s_1,s_2} = M_1^{1-2s_1-2s_2} \sum_{r=0}^n g^h_{r,s_1,s_2}(w)\,
   \bra{\bar{\2}}k|\1]^r \,[\bar{\2}|k\ket{\1}^{n-r} .
\label{eq:ThreePointGeneralPolynomial}
\ee
These coefficients $g^h_{r,s_1,s_2}(w)$ are a refined version
of $c^h_{(i),s_1,s_2}(w)$ in \eqn{eq:ThreePointStripped};
the main difference between them is some degree of rescaling by $M_2/M_1$.
The polynomial degree $n$ above is related to the maximal number of terms:
\be
n + 1\,=\,
\left\{
\begin{aligned}
   2s_1+1 , & \\
   s_1+s_2-|h|+1, & \\
   2s_2+1 , &
\end{aligned}
\right. \quad
\begin{aligned}
   s_2-s_1 & \geq |h| , \\
   |s_2-s_1| & < |h| , \\
   s_1-s_2 & \geq |h| ,
\end{aligned}
\label{eq:MaxNumberCoeffs}
\ee
\end{subequations}
This number matches the counting in~\cite{Conde:2016izb}.
For completeness, the above formulae~\eqref{eq:ThreePointGeneral} already
include the result of the above analysis for the negative messenger helicity,
in which case we used the anti-chiral basis, $|k]$ and $|p_1\ket{k}$.

Interestingly, the coupling counting~\eqref{eq:MaxNumberCoeffs} obeys the bound
\be
\text{\# coeffs.} \,\leq\, 2\text{min}(s_1,s_2) + 1 .
\ee
For instance, there is only one term
for the case of the scalar massive incoming state $s_1=0$.
Indeed, the constraint~\eqref{eq:ObviousConstraint}
immediately implies $s_2>|h|$,
so we get a trivial polynomial of degree $n(0,s_2,h) = 0$.
In that case, the amplitude takes the form
\be
\bm{{\cal A}}(p_2,s_2|p_1,s_1=0;k,h) = g^{|h|}_{0,0,s_2}(w) M_1^{1-2s_2}
   \braket{\bar{\2} k}^{s_2-h} [\bar{\2} k]^{s_2+h} .
\label{eq:Amplitude3pt}
\ee
Note that we have now assumed parity and thus conflated the dimensionless
coupling coefficients $g^{\pm h}_{0,0,s_2}(w)$
into the single coupling $g^{\pm |h|}_{0,0,s_2}(w)$, which still depends
on the absolute helicity value of the messenger particle.\footnote{In
   the worldline formalism, the parity assumption is called
   ``electric-magnetic'' duality~\cite{Goldberger:2005cd,Porto:2007qi}.
}

\subsection{Minimal mass-changing amplitudes}
\label{sec:minimal3pt}

As a minor digression, let us note that, for non-zero initial spin,
the proliferation of possible effective couplings
in the mass-changing three-point amplitude~\eqref{eq:ThreePointGeneral}
may be reduced if we come up with some notion of minimality.
Indeed, in a similar situation in the equal-mass case, $M_1=M_2$,
Arkani-Hamed, Huang and Huang \cite{Arkani-Hamed:2017jhn}
managed to single out the so-called ``minimal'' amplitudes
by considering its massless limit.
For positive helicity, these minimal amplitudes include, for instance,
\be
{\cal A}(p_2,s|p_1,s;k,h) = g_0^h (p_1\!\cdot \varepsilon_k^+)^h
   \braket{\bar{\2} \1}^{2s} ,
\label{eq:AHHStripped}
\ee
where for simplicity we have assumed $s_1=s_2=s$.
In other words, the stripped amplitude is proportional to
the tensor product of ${\rm SL}(2,\mathbb{C})$ Levi-Civita tensors
$(\epsilon^{2s})_{\{\alpha\},\{\beta\}}$.

To expose a similar unique structure in the unequal-mass case,
where the couplings correspond to the terms
in the polynomial~\eqref{eq:TensorPolynomial3pt},
we may change the basis inside of it to the antisymmetric and symmetric combinations of the basis spinors:
\be
[\la \mu \oplus \mu \la]^n_{\{\alpha\},\{\beta\}}
 = [\epsilon \oplus \sigma]^n_{\{\alpha\},\{\beta\}} , \qquad
\epsilon_{\alpha\beta}
 = \frac{\la_\alpha \mu_\beta - \mu_\alpha \la_\beta}{\braket{\la\mu}} , \qquad
\sigma_{\alpha\beta}
:= \la_\alpha \mu_\beta + \mu_\alpha \la_\beta .
\ee
Since of course
$\bra{\1}^\alpha \bra{\bar{\2}}^\beta \epsilon_{\alpha\beta} = \braket{\1 \bar{\2}}$
and the symmetric combination leads to
\be
\bra{\1}^\alpha \bra{\bar{\2}}^\beta \sigma_{\alpha\beta}
 = \frac{M_2^2+M_1^2}{M_1^2} \braket{\1 \bar{\2}}
 + \frac{2M_2}{M_1} [\1 \bar{\2}] ,
\label{SymSpinorCombination}
\ee
the main amplitude factor can simply be expanded in the angle and square brackets:
\be
\bm{F}^h_{s_1,s_2} = M_1^{1-2s_1-2s_2+n} \sum_{r=0}^n
   \tilde{g}^h_{r,s_1,s_2}(w)\,
   \braket{\bar{\2} \1}^{n-r} [\bar{\2}\1]^r .
\label{eq:ThreePointGeneralPolynomialAlt}
\ee
So we propose to define the minimal mass-changing stripped amplitudes
as those with highest power in $\epsilon_{\alpha\beta}$, or, equivalently,
\begin{align}
\label{eq:ThreePointPlusAmpMin}
 & {\cal A}_\text{min}(p_2,s_2|p_1,s_1;k,h)\\ &
 = \tilde{g}^{h}_{0,s_1,s_2}(w) 
\left\{
\begin{aligned}
 & M_1^{1-2s_2} \braket{\bar{\2} \1}^{2s_1}
   \braket{\bar{\2}k}^{s_2-s_1-h} [\bar{\2}k]^{s_2-s_1+h}\!, \\
 & M_1^{1-s_1-s_2-h} \braket{\bar{\2} \1}^{s_1+s_2-h}
   [\bar{\2}k]^{s_2-s_1+h} [k\1]^{s_1-s_2+h}\!, \\
 & M_1^{1-2s_2} \braket{\bar{\2} \1}^{2s_2}
   \braket{k \1}^{s_1-s_2-h} [k \1]^{s_1-s_2+h}\!,
\end{aligned}
\right. \quad
\begin{aligned}
   s_2-s_1 & \geq h \geq 0 ,\\
  |s_2-s_1| & < h , \\
   s_1-s_2 & \geq h \geq 0 .
\end{aligned} \nn
\end{align}
It is clear that for $s_1=0$ and $s_2>|h|$, the minimal-coupling amplitude
coincides with the previously defined amplitude~\eqref{eq:Amplitude3pt}.
Moreover, let us note in passing that these amplitudes satisfy
the double-copy prescription explored in the presence of
massive spinning states in~\cite{Johansson:2019dnu,Bautista:2019evw}.

We hope to explore these amplitudes in more detail elsewhere,
whereas in the rest of this paper for the sake of simplicity
we focus on the mass-changing amplitudes~\eqref{eq:Amplitude3pt}
with the non-spinning initial state,
which we use to model the radiation absorption by a Schwarzschild black hole.
In this context, it is important to note that
if we assume locality of the EFT Lagrangian that implies the above amplitudes,
the dimensionless coupling constants $g^h_{0,s_1,s_2}(w)$
may then be constrained to only have non-negative powers of~$w$.
Unfortunately, a rigorous proof of this statement may be to technical
and require dealing with all sorts of field redefinitions.
So for the purposes of this paper,
let us simply impose that $g^h_{0,0,s_2}(w)$ have no poles in $w$:
\be\!
g^h_{0,0,s_2}(w) = {\cal O}(w^0) \qquad\!\Rightarrow\!\qquad
\bm{{\cal A}}(p_2,s_2|p_1,s_1=0;k,h) = {\cal O}(w^{s_2}) ,\!\!\qquad w \to 0,
\label{eq:wBehavior}
\ee
which constitutes is a non-trivial EFT modeling assumption.

\section{Absorption from mass-changing amplitudes}
\label{sec:Absorption3pts}

In this section we combine the ingredients from the previous two sections:
the partial-wave absorption setup leading to the matrix element~\eqref{eq:SphHelSMatrixCov},
and the mass-changing three-point amplitudes~\eqref{eq:Amplitude3pt}.
The goal of this section will be to derive the absorptive cross-section
as a function of the effective coupling coefficients~$g^{|h|}_{0,0,s_2}(w)$.

\subsection{Mass-changing amplitudes as harmonics}
\label{sec:SpinConservation}

Focusing on the mass-changing amplitudes~\eqref{eq:Amplitude3pt},
it is rewarding to notice that they are simply proportional to the spin-weighted spherical harmonics~\eqref{SphHarmSpinors4d}, namely
\be
{\cal A}_{\scriptsize\underbrace{1 \dots 1}_{s_2-m}
	\underbrace{2 \dots 2}_{s_2+m}}(p_2,s_2|p_1;k,h)\!
 = M_1 g_{0,0,s_2}^{|h|}\!(w) (-1)^{s_2-h} w^{s_2}
   {}_h\tilde{Y}_{s_2,m}(k;u_2)\!
 =:{\cal A}_{s_2,m}^h(p_2|p_1;k) .
\label{eq:AmplitudeAsHarmonic2}
\ee
However, the harmonics are defined with respect to $u_2^\mu$,
which is counterproductive for plugging these amplitudes in the partial-wave absorption formula~\eqref{eq:SphHelSMatrixCov},
where $u_1^\mu$ is fixed but $u_2^\mu$ changes along with the integration variable~$k^\mu$.
So we wish to make the transition between the two velocity vectors,
which are related by the boost
\be
u^\rho_2  = L^\rho_\sigma(u_2\leftarrow u_1) u_1^\sigma
 = \exp\!\Big( \tfrac{i \log(u_1 \cdot u_2 + \sqrt{(u_1 \cdot u_2)^2-1})}{\sqrt{(u_1 \cdot u_2)^2-1}}
	u_1^\mu u_2^\nu \Sigma_{\mu\nu}\!\Big)^\rho_{~\sigma} u_1^\sigma .
\ee
The corresponding spinor transformations, given by \eqn{eq:SpinorTransform},
may be rewritten as
\be\!\!\!
\ket{u_2^a} = \frac{\sqrt{M_1}}{\sqrt{M_2}} 
\bigg( \ket{u_1^a} + \frac{|k|u_1^a] }{M_1\!+\!M_2} \bigg) , \qquad
|u_2^a] = \frac{\sqrt{M_1}}{\sqrt{M_2}} 
\bigg( |u_1^a] + \frac{|k\ket{u^a}}{M_1\!+\!M_2} \bigg) ,
\label{eq:SpinorTransform2}
\ee
where we have used
$\mu := u_1 \cdot u_2 + \sqrt{(u_1 \cdot u_2)^2-1} = M_2/M_1$.
The net effect of this is that the projection of the massive spinors
onto the directions $\ket{k}$ and $|k]$ is invariant under this boost,
so the spherical harmonics are related simply by
\be
\braket{2^a k} = \braket{1^a k} , \qquad 
[2^a k] = [1^a k]\qquad \Rightarrow\qquad {}_h\tilde{Y}_{s_2,m}(k;u_2)
= {}_h\tilde{Y}_{s_2,m}(k;u_1) .
\ee
(This is because we switch between rest frames of $p_1$ and $p_2=p_1+k$ inside the harmonics in the same direction~$k$.)
The caveat here is that the spin of particle~$2$ is now quantized along
$L^\mu{}_\nu(u_2\leftarrow u_1) n_1^\sigma$,
\ie the boost of the spin quantization axis of particle~$1$,
which may be arbitrary but has to be the same for every $p_2=p_1+k$.
With this restriction in mind, we may rewrite the three-point amplitude as
\be
{\cal A}_{s_2,m}^h(p_2|p_1;k)
= M_1\,g_{0,0,s_2}^{|h|}(w)(-1)^{s_2-h} w^{s_2}\,{}_h\tilde{Y}_{s_2,m}(k;u_1) .
\label{eq:AmplitudeAsHarmonic1}
\ee

Let us now introduce the spherical scattering amplitude\footnote{Note that
   the definition~\eqref{eq:SphHelAmplitudeCov} ignores
   the delta function $\hat{\delta}^4(p_1+k-p_2)$,
   which accompanies the scattering amplitude and imposes momentum conservation.
   Although it will play a role in the cross-section calculation in the next section,
   the above definition can still be found useful.
}
\be
\label{eq:SphHelAmplitudeCov}\!\!
{\cal A}_{\{b\}}(p_2,s_2|p_1;\omega,j,m,h)\!
:=\!\frac{4\pi}{\sqrt{2\omega}}\!\int_k\!
   \hat{\delta}(k \cdot u_1 -\omega)\,{}_{-h}Y_{j,m}(k;u_1)
   {\cal A}_{\{b\}}(p_2,s_2|p_1;k,h)\!
\ee
in an analogous manner to \eqn{eq:SphHel2PlaneWaveCov}.
Using the conjugation and orthogonality properties
\eqref{SphHarmSpinors4dConj} and~\eqref{SphHarmCovNorm},
we find
\begin{align}
{\cal A}_{\scriptsize\underbrace{1 \dots 1}_{s_2-m'}
   \underbrace{2 \dots 2}_{s_2+m'}}(p_2,s_2|p_1;\omega,j,m,h) \nn
 = \frac{(-1)^{-2j+m+h}\!}{\pi\sqrt{2\omega}}\!\int\!d^4 k\;\!
   \delta^+(k^2) \delta(k \cdot u_1 - \omega) & \\ \times\,
   {}_hY_{j,-m}^*(k;u_1)
   {\cal A}_{s_2,m'}^h(p_2|p_1;k) &
\label{eq:SphHelAmplitudeCovMin} \\
 = (-1)^{-j} \delta^j_{s_2} \delta^{-m}_{m'} M_1^{3/2}
   \Big[ 4\pi (2j\!+\!1) {\textstyle {2j \choose j+m}}
          {\textstyle {2j \choose j+h}} \Big]^{-1/2}
   g_{0,0,j}^{|h|}(w)\,w^{j+1/2} & \big|_{w=2\omega/M_1} . \nn
\end{align}
This neatly expresses the angular-momentum conservation law.
This simple form of the spherical scattering amplitude is valid
under our assumption that the magnetic quantum number~$m'$ is defined
with respect to the axis $L^\mu{}_\nu(u_2\leftarrow u_1) n_1^\sigma$.

\subsection{Leading-order absorption cross-section}
\label{sec:LOabsXsection}

We are now ready to construct the leading absorption cross-section
from the three-point amplitudes discussed above.
The inclusive cross-section for the spherical scattering setup
described in \sec{sec:SphHelicity} is
\cite{Goldberger:2005cd,Endlich:2016jgc}
\be
\sigma_\text{inc}(\omega_\text{cl},j,m,h) 
= \frac{\pi}{\omega_\text{cl}^2} P_\text{inc}(\omega_\text{cl},j,m,h)
= \frac{\pi}{\omega_\text{cl}^2} \sum_X
\frac{\big|\braket{X|S|\psi;\gamma}\big|^2}
{\braket{X|X} \braket{\psi|\psi} \braket{\gamma|\gamma}} ,
\label{eq:Prob2Xsection}
\ee
It is invariant under the basis choice for the outgoing states.
The leading contribution due to absorption is then
given by the 3-point process:
\be
P_\text{inc}^\text{LO}(\omega_\text{cl},j,m,h)
 = V\!\int_0^\infty\!\!dM_2^2 \rho(M_2^2) \int\!\hat{d}^3p_2
   \frac{\big|\braket{p_2|S|\psi;\gamma}\big|^2}
        {\braket{p_2|p_2} \braket{\psi|\psi} \braket{\gamma|\gamma}} .
\label{eq:SphHelProb}
\ee
Here $V:=\braket{p_2|p_2}/(2p_2^0) = \hat{\delta}^3(\bm{0})$
is the space volume, which immediately cancels
against the normalization of the outgoing state,
for which we have temporarily suppressed any quantized degrees of freedom.
We have also been compelled to include the spectral density
$\rho(M_2^2)$, which is positive and normalized to 1:
\be
\rho(q^2) \geq 0, \qquad \quad \int_0^\infty\!\!\rho(q^2)dq^2=1 .
\label{eq:SpectralDensityNorm}
\ee

In a conservative scenario, one may simply assume
$\rho(q^2) = \delta(q^2-M_1^2)$,
and the relevant amplitude would be the same-mass three-point amplitude.
More generally, it is allowed to contain suitably normalized delta-functions
for the ``elementary'' particles and the continuous part
due to multi-particle states.
Since we are interested in modeling absorption effects,
we are led to explore the continuous part of the spectrum for $q^2 > M_1^2$.
It can be checked that without a continuous part of the spectral-density function
the three-point kinematics would be overconstrained,
and the cross-section integration would yield a distribution.

In view of the normalization of the initial states,
$\braket{\psi|\psi} = \braket{\lambda|\lambda} = 1$,
the resulting leading-order probability is given by
\be
P_\text{inc}^\text{LO}(\omega_\text{cl},j,m,h)
 = \sum_{s_2}\!\int\!dM_2^2 \rho_{s_2}(M_2^2)
   \int_{p_2} \sum_{b_1,\dots,b_{s_2}}\!
   \big|\braket{p_2,s_2,\{b\}|S|\psi;\gamma}\big|^2 ,
\label{eq:SphHelProb2}
\ee
where we have now made the spin degrees of freedom of the outgoing state
explicit.
The integration over masses of $p_2$ different from $M_1$ is
what allows the three-point amplitude to exist on real kinematics
and thus makes this cross-section meaningful.
As we will see, momentum conservation will later fix this mass to
\be
M_2^2 = M_1^2 + 2M_1 \omega_\text{cl} .
\label{eq:M2}
\ee
After restoring $\hbar$ in front of $\omega_\text{cl}$,
it actually becomes sent back to $M_1$ in the classical limit,
so the spectral density will only by probed in the vicinity
of the original BH mass.
This, however, does not negate the crucial roles that
the unequal masses and the spectral density play
in allowing for a non-singular construction
of the cross-section from three-point amplitudes.

Coming back to the squared amplitude in the integrand of \eqn{eq:SphHelProb2},
we have
\beal\!\!
\sum_{\{b\}} \big|\braket{p_2,s_2,&\{b\}|S|\psi;\gamma}\big|^2
 = 8\pi^2\!\int_0^\infty\!
   \frac{\hat{d}\omega \hat{d}\omega'}{\sqrt{\omega \omega'}}
   \gamma_\zeta^*(\omega) \gamma_\zeta(\omega')\!
   \int_{p_1,p_1',k,k'}\!\!\psi_\xi^*(p_1) \psi_\xi(p_1') \\ \times\,&
   \hat{\delta}(k \cdot u_1-\omega) \hat{\delta}(k'\!\cdot u_1-\omega')
   \hat{\delta}^{4}(p_1+k-p_2) \hat{\delta}^{4}(p_1'+k'-p_2) \\ \times\,&
   {}_{-h}Y_{j,m}^*(k;u_1)\,{}_{-h}Y_{j,m}(k';u_1)\,
   {\cal A}^{*\{b\}}(p_2,s_2|p_1;k,h)\,
   {\cal A}_{\{b\}}(p_2,s_2|p_1';k'\!,h) ,\!\!\!\!
\label{eq:SphHelProb2A}
\eeal
where the summation over the little-group indices $\{b\}$ is now implicit.
We may use $\hat{\delta}^{4}(p_1+k-p_2)$ to perform the integration over $p_2$,
which leaves the on-shell constraint $\hat{\delta}((p_1+k)^2-M_2^2)$.
We then change the integration variables to
\be
p_\text{a}^\mu  := (p_1^\mu + p_1'^\mu)/2, \qquad \quad
q^\mu  := p_1'^\mu - p_1^\mu ,
\ee
and remove $q$ with $\hat{\delta}^4(q+k'-k)$
originating from $\hat{\delta}^4(p_1'+k'-p_2)$.
Thus we get
\beal
 & P_\text{inc}^\text{LO}(\omega_\text{cl},j,m,h)
 = 8\pi^2 \sum_{s_2}\!\int\!dM^2_2\rho_{s_2}(M_2^2)
   \int_0^\infty\!\frac{\hat{d}\omega \hat{d}\omega'}{\sqrt{\omega\omega'}}
   \gamma_\zeta^*(\omega) \gamma_\zeta(\omega')\!
   \int_{k,k'}\!\!\hat{\delta}(k \cdot u_1-\omega) \\ & \times
   \hat{\delta}(k'\!\cdot u_1-\omega')\,
   {}_{-h}Y_{j,m}^*(k;u_1)\,{}_{-h}Y_{j,m}(k';u_1)\!
   \int\!\hat{d}^4p_\text{a}\,
   \hat{\delta}^+(p_\text{a}^2 - M_1^2 - k'\!\cdot k/2)
   |\psi_\xi(p_\text{a})|^2 \\ & \times
   \hat{\delta}(2p_\text{a}\!\cdot k -  2p_\text{a}\!\cdot k')
   \hat{\delta}(M_1^2 + 2p_\text{a}\!\cdot k + k'\!\cdot k - M_2^2)\, 
   {\cal A}^{*\{b\}}(p_\text{a}\!+\!\tfrac{k+k'\!}{2},s_2|
                     p_\text{a}\!+\!\tfrac{k'-k}{2};k,h)\, \\ & \qquad
   \qquad \qquad \qquad \qquad \qquad \qquad \qquad \qquad \qquad~\;\times
   {\cal A}_{\{b\}}(p_\text{a}\!+\!\tfrac{k+k'\!}{2},s_2|
                    p_\text{a}\!+\!\tfrac{k-k'\!}{2};k',h) ,
\label{eq:SphHelProb3}
\eeal
where we have also used the convenient property
$\psi_\xi^*(p_\text{a}\!-\!\tfrac{q}{2}) \psi_\xi(p_\text{a}\!+\!\tfrac{q}{2}) 
= |\psi_\xi(p_\text{a})|^2$
of the momentum wavepackets~\eqref{eq:MomentumWF}.

\subsection{Absorption cross-section in classical limit}
\label{sec:LOabsXsectionCl}

So far no classical limit was taken,
and \eqn{eq:SphHelProb3} still represents a quantum probability.
To rectify that, we send $\xi \rightarrow 0$ and
evaluate the integral over $p_\text{a}$,
which in the presence of the squared wavefunction $|\psi_\xi(p_\rma)|^2$
and the mass-shell delta function has the effect of setting the momentum
$p_\text{a}^\mu$ to its classical value
$u_1^\mu \sqrt{M_1^2 + k'\!\cdot k/2} =: M_\text{a} u_1^\mu$.
Subsequently, using the delta function
$\hat{\delta}(2p_\text{a}\!\cdot k -  2p_\text{a}\!\cdot k')$
becomes $\hat{\delta}(\omega-\omega')/(2M_\text{a})$,
which removes the integration over $\omega'$.
In the integral over the remaining $\omega$, we send $\zeta \rightarrow 0$,
so the squared wavefunction $|\gamma_\zeta(\omega)|^2$
localizes it at the classical value~$\omega_{\rm cl}$.
In this way, the above probability becomes
\begin{align}\!\!\!
\lim_{\zeta \rightarrow 0} \lim_{\xi \rightarrow 0}
   P_\text{inc}^\text{LO}(\omega_\text{cl},j,m,h)
 = \frac{16\pi^3\!}{\omega_\text{cl}}
   \sum_{s_2}\!\int_{k,k'}\!\frac{1}{2M_\text{a}}
   \rho_{s_2}(M_1^2 + 2M_\text{a} \omega_\text{cl} + k'\!\cdot k)
   \hat{\delta}(k & \cdot u_1-\omega_\text{cl}) \nn \\ \times\,
   \hat{\delta}(k'\!\cdot u_1-\omega_\text{cl})\,
   {}_{-h}Y_{j,m}^*(k;u_1)\,{}_{-h}Y_{j,m}(k';u_1)\,
   {\cal A}^{*\{b\}}(p_\text{a}\!+\!\tfrac{k+k'\!}{2},s_2|
                     p_\text{a}\!+\!\tfrac{k'-k}{2};k &,h)\,
\label{eq:SphHelProb4}\\ \times
   {\cal A}_{\{b\}}(p_\text{a}\!+\!\tfrac{k+k'\!}{2},s_2|
                    p_\text{a}\!+\!\tfrac{k-k'\!}{2};k'& ,h)
   \big|_{p_\text{a} = M_\text{a} u_1}\!, \nn
\end{align}
where we have also taken the integral over $M_2^2$
using $\hat{\delta}(M_1^2 + 2M_\text{a} \omega + k'\!\cdot k - M_2^2)$.

Even though we have simplified the probability expression considerably,
the integrals over $k^\mu$ and $k'^\mu$ are still intertwined,
in particular because the spectral density and $M_\text{a}$
both depend on $k \cdot k'$.
Note, however, that the two massless momenta are constrained
to have the energy projection $\omega_\text{cl}$,
so $|k \cdot k'| \le 2\omega_\text{cl}^2$,
as most easily seen in the rest frame of $u_1^\mu$.
The basic classical-limit assumption $\omega_\text{cl} \ll M_1$ then implies
\be
|k^\mu|, |k'^\mu| ~\ll~ M_1 \qquad \Rightarrow \qquad
|k\cdot k'| ~\ll~ M_1 u_1\!\cdot k = M_1 u_1\!\cdot k' = M_1 \omega_\text{cl} .
\label{eq:kPrimedkExpansion}
\ee
Therefore, we may define the classical limit of the above probability as
\beal\!\!\!
P_\text{inc,\,cl}^\text{LO}
 = \frac{8\pi^3\!}{M_1\omega_\text{cl}} 
   \sum_{s_2} \rho_{s_2}(M_1^2)\! 
   \int_{k,k'}\!\!\hat{\delta}(k\!\cdot\!u_1-\omega_\text{cl})\,&
   {}_{-h}Y_{j,m}^*(k;u_1)\,{\cal A}^{*\{b\}}(p_2,s_2|p_1;k,h) \\ \times\,
   \hat{\delta}(k'\!\!\cdot\!u_1-\omega_\text{cl})\,&
   {}_{-h}Y_{j,m}(k';u_1)\,{\cal A}_{\{b\}}(p_2,s_2|p_1';k',h) ,\!\!\! \label{eq:SphHelProbCl}
\eeal
where for brevity we have now used the momenta
\be
p_1 = M_1 u_1+\tfrac{k'-k}{2} , \qquad
p_1' = M_1 u_1+\tfrac{k-k'\!}{2} , \qquad
p_2 = M_1 u_1+\tfrac{k+k'\!}{2} =: M_2 u_2
\label{eq:MomentaCl}
\ee
not as independent integration variables but to denote their classical values.
Note that in the expression above,
we have already assumed that the outgoing states
are described by a sufficiently smooth spectral-density function,
which makes sense because our EFT is meant to describe absorption of classical waves of arbitrary frequency (provided it is small).
Therefore, $\rho_{s_2}$ can be expanded in $\omega_\text{cl}/M_1$,
for which $2M_1\omega_\text{cl}$ and $k'\!\cdot k$ provide
linear and quadratic terms, respectively,
and both may be dropped,
leaving only the leading term $\rho_{s_2}(M_1^2)$ in the classical limit.

Let us now deal with the momentum dependence of the amplitudes,
which, as we have noticed in \eqn{eq:AmplitudeAsHarmonic2},
are proportional to the covariant spin-weighted spherical harmonics
${}_h\tilde{Y}_{s_2,m'}(k;u_2)$,
while their prefactors depend on the dimensionless ratio
\be
w := \frac{2p_1\!\cdot k}{M_1^2}
   ~\simeq~ \frac{2\omega_\text{cl}}{M_1}
   ~\simeq~ \frac{2p_1'\!\cdot k'}{M_1^2} =: w' .
\ee
Moreover, just as we did in \sec{sec:SpinConservation},
we may boost the time direction $u_2^\mu$ of either harmonic
to our preferred $u_1^\mu$,
with their difference now being equal to $(k+k')^\mu/2$,
but the result still being\footnote{More explicitly, one may use
the most general spinor transformations~\eqref{eq:SpinorTransformFull}
to observe:
\be
\ket{2^b}\!=\!U^b{}_a(u_2\!\leftarrow\!u_1) \sqrt{\!M_1\!}\;\!
\bigg\{\!\ket{u_1^a} + \frac{|k\!+\!k'|u_1^a] }{2(M_1\!+\!M_2)}\!\bigg\}
\,\Rightarrow\,
\braket{2^b k}\!=\!U^b{}_a(u_2\!\leftarrow\!u_1) \sqrt{\!M_1\!}\;\!
   \Big\{\!\braket{u_1^a k}
    + {\cal O}\big([u_1^a k'] \tfrac{\!\braket{k'k}\!}{M_1}\big)\!\Big\} , \nn
\label{eq:SpinorTransform3}
\ee
and similarly for the anti-chiral spinors.
We have thus exposed the spinorial (square-root) version
of the classical hierarchy assumption~\eqref{eq:kPrimedkExpansion}.
Coming back to the original three-point amplitude~\eqref{eq:Amplitude3pt},
one can then expose its classically meaningful term as (proportional to)
\be
\braket{2^b k}^{\odot (s_2-h)} \odot [2^b k]^{\odot(s_2+h)}
~\simeq~ M_1^{s_2} \big\{(U^b{}_a(u_2\!\leftarrow\!u_1)\big\}^{\odot 2s_2}
   \braket{u_1^a k}^{\odot (s_2-h)} \odot [u_1^a k]^{\odot(s_2+h)} , \nn
\ee
Moreover, the unitarity of the ${\rm SU}(2)$ transformation matrices
$U^b{}_a(u_2\!\leftarrow\!u_1)$ ensures that they cancel
in all inclusive-probability expressions, such as \eqn{eq:AmplitudeSquared},
and hence justifies our liberal treatment of the little-group indices,
also phrased as the quantization-axis choice assumption
in \sec{sec:SpinConservation}.
}
${}_h\tilde{Y}_{s_2,m'}(k;u_2) \simeq {}_h\tilde{Y}_{s_2,m'}(k;u_1)$.
Therefore, the squared amplitude is
\beal
{\cal A}^{*\{b\}}(& p_2,s_2|p_1;k,h) {\cal A}_{\{b\}}(p_2,s_2|p_1';k',h) \\ &
   \simeq M_1^2\,|g_{0,0,s_2}^{|h|}(w)|^2 w^{2s_2}\!\!
   \sum_{m'=-s_2}^{s_2}\!\!\!{\textstyle {2s_2 \choose s_2+m'}}\,
   {}_h\tilde{Y}_{s_2,m'}^*(k;u_1)\,{}_h\tilde{Y}_{s_2,m'}(k';u_1) .
\label{eq:AmplitudeSquared}
\eeal
Having thus completely disentangled the integrations in $k$ and $k'$,
we may evaluate
\begin{align}
\label{eq:SphHelProbCl2}
P_\text{inc,\,cl}^\text{LO}(\omega_\text{cl},j,m,h)
=\,&\frac{8\pi^3\!}{\omega_\text{cl}} M_1 \sum_{s_2} \rho_{s_2}(M_1^2)
   |g_{0,0,s_2}^{|h|}(w)|^2  w^{2s_2} \\ \times\!\!
   \sum_{m'=-s_2}^{s_2}\!\!{2s_2 \choose s_2+m'}\, &
   \bigg| \int\!\hat{d}^4k\,\hat{\delta}^+(k^2)
          \hat{\delta}(k \cdot u_1-\omega_\text{cl})\,
          {}_{-h}Y_{j,m}(k;u_1)\,{}_h\tilde{Y}_{s_2,m'}(k;u_1)
   \bigg|^2 \nn \\
=\,&\frac{M_1^2}{4(2j+1)} {2j \choose j+h}^{\!-1}
   \rho_j(M_1^2)\,|g_{0,0,j}^{|h|}(w)|^2\,w^{2j+1} , \nn
\end{align}
where we have used the conjugation and orthogonality properties
\eqref{SphHarmSpinors4dConj} and~\eqref{SphHarmCovNorm}.\footnote{Alternatively, the result of \eqn{eq:SphHelProbCl2}
may be obtained by plugging in the previously computed
spherical scattering amplitudes~\eqref{eq:SphHelAmplitudeCovMin}
with classical momentum values~\eqref{eq:MomentaCl}:
\be
P_\text{inc,\,cl}^\text{LO}(\omega_\text{cl},j,m,h)
 = \frac{\pi}{M_1} \rho_{s_2}(M_1^2)\!
   \sum_{m'=-s_2}^{s_2}\!\!{\textstyle{2s_2 \choose s_2+m'}}
   \big| {\cal A}_{\scriptsize\underbrace{1 \dots 1}_{s_2-m'}
   \underbrace{2 \dots 2}_{s_2+m'}}(p_2,s_2|p_1;\omega_\text{cl},j,m,h) \big|^2 . \nn
\ee
}
Note that its power in $\omega_\text{cl} = M_1 w/2$ is dictated
by the total angular-momentum quantum number~$j$ of the incoming wave,
which also determines the absorptive three-point coupling that is being probed.

In this way, we have arrived at the partial-wave absorption cross-section
\beal\!
\sigma_\text{inc,\,cl}^\text{LO}(\omega_\text{cl},j,m,h)
:= &\,\frac{\pi}{\omega_\text{cl}^2} P_\text{inc,\,cl}^\text{LO}(\omega_\text{cl},j,m,h) \\
 = &\,\frac{\pi}{4\omega_\text{cl}^2} \frac{(j+h)!(j-h)!}{(2j+1)!}
   M_1^2\rho_j(M_1^2)\,|g_{0,0,j}^{|h|}(w)|^2\,w^{2j+1} ,
\label{eq:SphHelXsectionCovMain}
\eeal
where $w=2\omega_\text{cl}/M_1$.
We will deal with the apparent issue of $w$ being small
for $\omega_\text{cl} \ll M_1$ in the next section.

\section{Matching to microscopic calculation}
\label{sec:Matching}

The absorptive cross-section
of a Kerr black hole in general relativity was originally obtained by
Starobinsky, Churilov~\cite{Starobinsky:1973aij,Starobinskil:1974nkd}
and Page~\cite{Page:1976df,Page:1976ki} for the $j=|h|$ case
and recently generalized to arbitrary $j$
in~\cite{Chia:2020yla,Charalambous:2021mea,Ivanov:2022qqt}.
However, the dynamics of non-spinning BHs under small perturbations
dates back to Regge and Wheeler~\cite{Regge:1957td},
who proved linear stability of Schwarzschild BHs.
From the point of view of the EFT amplitudes,
which treat the BH as a particle,
the GR results serve as the microscopic computation,
to which the effective couplings should be matched.

\subsection{Classical absorption cross-section}
\label{sec:Classical}

In the general case of wave of spin~$|h|$
scattering off a spinning BH,
the transmission and scattering coefficients are usually obtained
by solving the Teukolsky equation
\cite{Teukolsky:1973ha,Press:1973zz,Teukolsky:1974yv}.
In this work, we focus on the simpler case of non-spinning BHs.
Let the Schwarzschild radius be $r_\text{S} := 2GM_1$
and $\omega$ the frequency of the classical spin-$|h|$ wave,
which obey $r_\text{S}\omega \ll 1$.
Then the absorption cross-section is given by
\cite{Ivanov:2022qqt}\footnote{We have dropped the prefactor $(2j+1)$
from the expressions in the literature, which comes from summing
over $m=-j,\dots,j$.
}
\be
\sigma_\text{abs}^\text{Schw}(\omega,j,m,h)
 = \frac{(-1)^h 2\pi}{\omega^2}
   \frac{(j+h)! (j-h)!}{(2j)! (2j+1)!}
   (2r_\text{S}\omega)^{2j+1} {\rm Im} F_{-hjh}^\text{Schw}(\omega) .
\label{eq:MicroscopicXsec}
\ee
Here $F_{hjm}^\text{Schw}$ is the harmonic near-zone response function
\be
F_{hjm}^\text{Schw}(\omega) = i(-1)^h\,r_\text{S} \omega\,
   \frac{(j+h)! (j-h)!}{(2j)! (2j+1)!}
   \prod_{l=1}^{j} \big[l^2\!+ (2r_\text{S}\omega)^2\big] ,
\label{eq:ResponseFunction}
\ee
which does not depend on the quantum number~$m$,
since we wrote it for a non-spinning black hole.
We have followed the GR literature \cite{Chia:2020yla,Ivanov:2022qqt}
in writing the cross-section~\eqref{eq:MicroscopicXsec}
using the response function so as to point out that
it is the latter that contains the expansion in $\omega$,
whereas the outside power of $\omega$ is fixed to be $2j-1$,
combined from the $\pi/\omega^2$ dimensionful prefactor
and $2j+1$ powers of a dimensionless frequency combination.
This factorization mimics the structure
of the corresponding EFT cross-section~\eqref{eq:SphHelXsectionCovMain},
that we are going to match to next.

Our focus, however, is on the leading powers in $\omega$ for each $j$,
which amounts to replacing the complicated product in the response function~\eqref{eq:ResponseFunction} by $(j!)^2$. We obtain
\be
\sigma_\text{abs,\,LO}^\text{Schw}(\omega,j,m,h) = 4\pi r_\text{S}^2
   \left[\frac{j! (j+h)! (j-h)!}{(2j)! (2j+1)!}\right]^2\!
   (2r_\text{S}\omega)^{2j} ,
\label{eq:MicroscopicXsecLO}
\ee
where of course $|m|, |h| \leq j$, and otherwise it vanishes.

\subsection{Scales and effective couplings}
\label{sec:ClassicalScales}

In order to properly compare the classical and EFT results,
it is helpful to restore $\hbar$ (while leaving $c=1$ throughout this paper).
This introduces the distinction between frequencies/lengths and masses:
\be
[\hbar] = L \times M , \quad
[M_1] = [\omega_\text{cl}] = M , \quad
[\omega] = L^{-1} , \quad
[r_\text{S}] = L , \quad
[G] = L \times M^{-1} ,
\label{eq:Dimensionalities}
\ee
where we have insisted on the new meaning of
$\omega := \omega_\text{cl}/\hbar$ as the wave frequency.
We should also multiply the right-hand side
of the cross-section given from~\eqref{eq:SphHelXsectionCovMain} by $\hbar^2$,
so as to switch its dimensionality from $M^{-2}$ to $L^2$.
This gives
\be
\sigma_\text{inc,\,cl}^\text{LO}(\omega,j,m,h)
 = \frac{\pi}{4\omega^2} \frac{(j+h)!(j-h)!}{(2j+1)!}
   M_1^2\rho_j(M_1^2)\,|g_{0,0,j}^{|h|}(\omega)|^2
   \bigg(\!\frac{2\hbar\omega}{M_1}\!\bigg)^{2j+1} .
\label{eq:MacroscopicXsec}
\ee
Here we have left the effective couplings $g_{0,0,s_2}(\omega)$
fully dimensionless.
Note, however, that in view of the presence of multiple scales,
they are now allowed to depend on $\omega$ through more than just
the ${\hbar\omega}/{M_1}$ ratio.

Let us discuss the two basic assumptions underlying
the EFT- and GR-based computations, \ie
$\hbar \omega \ll M_1$ and $r_\text{S} \omega \ll 1$.
The point is that the latter is a much stronger inequality than the former,
as the Schwarzschild radius must of course be assumed to be
many orders of magnitude larger than the Compton wavelength of the black hole:
\be
\omega ~\ll~ \frac{1}{r_\text{S}\!} ~\ll~
\frac{1}{\lambda_\text{C}\!} := \frac{M_1}{2\pi\hbar} .
\label{eq:ScaleHierarchy}
\ee
Otherwise, we would be in the realm of quantum gravity and not GR.
It is then clear that in the context of comparing the classical
and amplitude-based results, which both constitute frequency expansions,
we should retain only the leading order in ${\hbar\omega}/{M_1}$,
but classical frequency dependence may still be present
in the form of~$r_\text{S} \omega$.

Therefore, matching the leading-order
cross-sections~\eqref{eq:MicroscopicXsecLO} and~\eqref{eq:MacroscopicXsec}
directly, we obtain
\be
M_1^2\rho_j(M_1^2)\,|g_{0,0,j}^{|h|}(\omega)|^2
 = \frac{8 [j!]^2 (j+h)! (j-h)!}{[(2j)!]^2 (2j+1)!}
   \bigg(\!\frac{M_1 r_\text{S}}{\hbar}\!\bigg)^{\!2j+1} r_\text{S} \omega .
\label{eq:MatchingLO}
\ee
It is perhaps more aesthetically pleasing to rephrase this relationship
in terms of the classical response function:
\be
M_1^2\rho_j(M_1^2)\,|g_{0,0,j}^{|h|}(\omega)|^2
 = \frac{8 (-1)^h}{(2j)!}
   \bigg(\!\frac{M_1 r_\text{S}}{\hbar}\!\bigg)^{\!2j+1}
   {\rm Im} F_{-hjh,\,\text{LO}}^\text{Schw}(\omega) .
\label{eq:MatchingResponseFunction}
\ee
In other words, we have related the $j$-th effective absorptive coupling
squared to the imaginary part of the response function,
resembling a dispersion relation.
It might seem awkward to keep $\hbar$ in the now classically meaningful
cross-section expression~\eqref{eq:MacroscopicXsec},
as well as \eqns{eq:MatchingLO}{eq:MatchingResponseFunction}.
However, the effective couplings are a priori arbitrary,
and we are free to make convenient modeling assumptions about them,
so nothing prevents us from absorbing the Planck constants
into them as\footnote{Recalling the form of the three-point amplitude~\eqref{eq:Amplitude3pt},
we see that the effective-coupling rescaling~\eqref{eq:RescaleCouplings}
amounts to replacing massless momenta $k^\mu$
with wavevectors $\bar{k}^\mu := k^\mu/\hbar$,
which is commonplace in the KMOC formalism~\cite{Kosower:2018adc},
plus an additional overall $\hbar^{-1/2}$.
}
\be
\bar{g}_{0,0,s_2}^{|h|}(\omega) := \hbar^{s_2+1/2} g_{0,0,s_2}^{|h|}(\omega) .
\label{eq:RescaleCouplings}
\ee

Comparing the macroscopic and microscopic formulae~\eqref{eq:MicroscopicXsec}
and~\eqref{eq:MacroscopicXsec}, there are a number of things to observe.
\begin{itemize}[leftmargin=21pt,parsep=0pt]

\item
Both cross-sections are consistent in that neither depends on
the magnetic quantum number~$m$ of the spherical wave.

\item
The EFT cross-section~\eqref{eq:MacroscopicXsec} reproduces the static limit
$\sigma_\text{inc,cl}^\text{LO}(\omega\!=\!0,j,m,h) = 0$
for electromagnetism and gravity ($|h|=1$ and $2$, respectively)
because of the locality assumption~\eqref{eq:wBehavior}
that the Wilson coefficients have no negative powers of~$\omega$.
This can be considered as an EFT prediction,
\ie it holds prior to the matching of the three-point couplings.

\item
As previously mentioned,
the growth of the superficial leading power in $\omega$ with~$j$
is the same in both cross-sections, where by superficial we mean excluding
the $\omega$ dependence in the response function and the three-point couplings.
In other words, the matching~\eqref{eq:MatchingLO} contains
that same leading power of $\omega$ for any $j$,
and the cleaner matching~\eqref{eq:MatchingResponseFunction} between
the response functions and the three-point couplings
does not involve $\omega$ at all.

\item
In the EFT cross-section~\eqref{eq:MacroscopicXsec},
every three-point coupling $|g_{0,0,s_2}^{|h|}(\omega)|^2$
comes accompanied by the dimensionless combination $M_1^2\rho_{s_2}(M_1^2)$
involving the spectral density. Its appearance is very sensible
from the QFT point of view, as the probability that a massive particle
absorbs a lower-energy massless boson is necessarily proportional
to the number of possible resulting states with nearly identical mass.
However, since it always accompanies the couplings,
one may regard the complete expression
$M_1 \sqrt{\rho_{s_2}(M_1^2)}\,g_{0,0,s_2}^{|h|}(\omega)$
as a kind of effective coupling.
Alternatively, if one's focus is on modeling classical effects
that are guaranteed to be insensitive
to the difference between spectral densities for different masses and spins,
one could consider disregarding
the normalization constraint~\eqref{eq:SpectralDensityNorm} altogether
and make a modeling assumption $\rho_{s_2}(M_1^2) = 1/M_1^2$.

\item
Perhaps most importantly, we observe that the matching~\eqref{eq:MatchingLO}
means that
\be
g_{0,0,s_2}^{|h|}(\omega) = {\cal O}(G^{s_2+1}) ,
\label{eq:PMbehavior}
\ee
in the post-Minkowskian expansion, since $r_\text{S} = GM$.
In other words, the amplitude that the scalar particle
which models a Schwarzschild black hole absorbs
a spherical wave with total angular momentum~$j$
is a $(j\!+\!1)$-PM object.

\item
For gravity ($|h|=2$), the PM behavior~\eqref{eq:PMbehavior} means that
the Wilson coefficient starts at $s_2=2$ and scales as $\mathcal{O}(G^{3})$,
whereas the resulting leading absorption cross-section is at 6PM
for a $j=2$ spherical wave,
and higher harmonics are suppressed in the PM expansion.

\end{itemize}

In view of the classical cross-section~\eqref{eq:MicroscopicXsec} being
a polynomial in $\omega$ spanned by $\{\omega^{2j},\dots,\omega^{4j}\}$,
one might hope that higher orders in $r_\text{S}\omega$ could be retained,
as long as they are captured
by the response function~\eqref{eq:ResponseFunction}
in a perturbation scheme~\cite{Ivanov:2022qqt}
that is consistent classically.
Unfortunately, this is not the case in the present three-point setup,
because going to higher orders requires a more subtle matching.
Indeed, the higher orders in $r_\text{S}\omega$
in the EFT cross-sections~\eqref{eq:MacroscopicXsec}
are subject to interference from higher-multiplicity amplitudes.
More specifically, the next order in the cross-section is ${\cal O}(G^{2j+4})$,
for which the EFT treatment must, for instance, include amplitudes with
two additional conservative couplings to the graviton,
each ${\cal O}(\!\sqrt{G})$.
Furthermore, double-graviton absorption or even the mass-changing contact terms contribution to the Compton amplitude might contribute to this matching.
We will discuss these matters further below in \secs{sec:Diagrammatics}{sec:HigherPM}.

Generalizing this result to spinning objects is another story.
In the non-spinning case, the coupling constant $G$ only enters
in the Schwarzschild radius $r_\text{S}$,
whereas in the Kerr case where the dimensionless spin ratio
$a_* = a/GM$ also contains negative powers in $G$.
This shows that for Schwarzschild black holes,
the first contribution to such amplitudes is at 6PM
(as can be reproduced by off-shell EFT methods~\cite{Goldberger:2005cd,Porto:2007qi}),
while it comes at a lower order for Kerr black holes
due to the negative power of $G$ in~$a_*$.
For instance, the authors of~\cite{Saketh:2022xjb} consider four-point contact interactions,
where such effects come at spin-5 in ${\cal O}(G)$ amplitudes.
Nevertheless, the general formalism presented in this paper
does allow to go to higher orders in spin,
and we leave this for future work.

In this purely on-shell approach, we have modeled the absorption effects by allowing a changing-mass amplitude from $s_1= 0$ to a spinning degree of freedom and the leading order corresponds to a $s_2=2$ particle.
We have observed some similarities with the worldline EFT approach
\cite{Goldberger:2005cd,Porto:2007qi,Goldberger:2017ogt},
where the point-particle action coupled to the Weyl tensor is not enough
to model absorption. One then has to introduce electric and magnetic composite operators $Q_{ab}^E$ and $Q_{ab}^B$ representing new degrees of freedom,
which carry two indices and couple to electric and magnetic components
of the Weyl tensor $E^{ab}$ and $B^{ab}$, respectively.
While in our approach higher orders require considering $s_2\geq 2$ particles and higher-multiplicity amplitudes, on the worldline
higher-derivative operators acting on the Weyl tensor
and multi-index composite operators
are needed to improve the calculation beyond $\omega^4$,
which is explored \eg in~\cite{Saketh:2022xjb}.

\section{Coherent-state cross-section}
\label{sec:CoherentXsec}

A proper description of the interaction between a gravitational wave
and a compact object using scattering amplitudes requires the use
of a coherent-state formalism to model the incoming and outgoing wave
\cite{Endlich:2016jgc,Ilderton:2017xbj,Cristofoli:2021vyo}.
In \sec{sec:Mechanics}, we have circumvented it
by using a single-graviton state with a wavefunction
peaked at the classical frequency~$\omega_{\rm cl}$.
The point of this section is two-fold:
\begin{itemize}[leftmargin=21pt,parsep=0pt]
\item
substantiate the leading-order calculation via the coherent-state framework,
\item
explain how higher-order calculations may be done in a similar fashion.
\end{itemize}
We focus on a coherent cross-section-based (or probability-based) formalism instead of an observable-based one~\cite{Kosower:2018adc}.
We start with a quantum description and make gradual assumptions
relevant to the classical limit.

\subsection{Elastic-inelastic separation}
\label{sec:Separation}

\begin{figure}[t]
\centering
\vspace{-15pt}
\includegraphics[width=7cm]{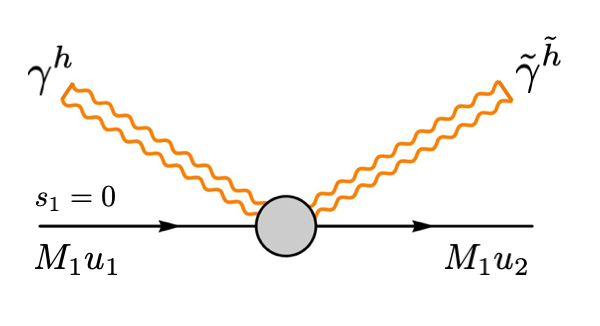}
\includegraphics[width=7cm]{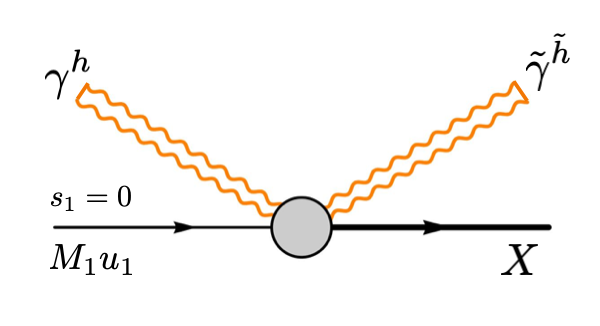}
\vspace{-20pt}
\caption{Gravitational diagrams in a non-spinning black-hole-wave interaction}
\label{fig:coherent_X}
\end{figure}

The initial state for our absorption process
consists of a heavy non-spinning particle
$\ket{\psi_1}$ and a wave of helicity $h$
modeled by a massless coherent state $\ket{\gamma^h}$.
We write
\be
\ket{\rin} := \ket{\psi_1 ;\gamma^h}
 =\int_{p_1} \psi_{\xi}(p_1) e^{ib\cdot p_1/\hbar} \ket{p_1;\gamma^h} ,
\label{eq:InState}
\ee
where the relativistic momentum-space wavefunction $\psi_\xi(p_1)$
peaks at the classical momenta $p_{1,\text{cl}}^\mu = M_1 u^\mu_1$,
as discussed in \sec{sec:Mechanics}.
We have also allowed for an impact parameter.
For the final state, we should distinguish two cases:
\begin{itemize}
\item[\text{(c)}]
a different coherent state $\ket{\tilde{\gamma}^{\tilde{h}}}$,
but the heavy particle's mass is preserved;
\item[\text{(nc)}]
a different coherent state $\ket{\tilde{\gamma}^{\tilde{h}}}$
and an unspecified particle $\ket{X}$ with $M_2 \neq M_1$.
\end{itemize}
The two cases are depicted in \fig{fig:coherent_X},
and we need to integrate over the possible final states.
Despite these assumptions,
the formalism easily allows for initial spinning states,
and we delay the specification of the massless coherent-state type
(plane-wave or partial-wave) to later on.
It is also worth commenting that even though case~$\text{(c)}$ has the same mass as the initial state, intermediate mass transitions are allowed (\eg Compton scattering with different masses in the factorization channels).

The need to separate these two cases on the quantum side
comes from the discontinuous nature of basic
scattering-amplitude building blocks at $M_2=M_1$,
as discussed in \sec{sec:3pts},
and on the classical side from the usual separation
between conservative and non-conservative effects.
The total probability will then include the following mass-preserving and mass-changing probabilities
\be
P_{\gamma \to \tilde{\gamma}}
 = P^\text{(c)}_{\gamma \to \tilde{\gamma}}
 + P^\text{(nc)}_{\gamma \to \tilde{\gamma}} .
\ee
For the first one, we may write
\beal
P^\text{(c)}_{\gamma \to \tilde{\gamma}} &
 = \sum_{2s_2=0}^\infty\,\sum_{b_1,\dots,b_{2s_2}=1,2}\,
   \int_{p_2}\!\bra{\rin} S^\dagger
   \ket{p_2,s_2,\{b\};\tilde{\gamma}^{\tilde{h}}}
   \bra{p_2,s_2,\{b\};\tilde{\gamma}^{\tilde{h}}} S \ket{\rin} .
\label{eq:ProbCons}
\eeal

In the second case, we are interested in the probability of
all different configurations $X$ involving a heavy particle of mass
$M_2 \neq M_1$:
\be
P^\text{(nc)}_{\gamma \to \tilde{\gamma}}
 = \sum_{X \ni M_2 \neq M_1}\!\!|\bra{X; \tilde{\gamma}^{\tilde{h}}} S \ket{\rin}|^2
 = \sum_{X \ni M_2 \neq M_1}\!\!\bra{\rin} S^\dagger
   \ket{X; \tilde{\gamma}^{\tilde{h}}}
   \bra{X; \tilde{\gamma}^{\tilde{h}}} S \ket{\rin}.
\ee
The crucial point now is to determine
what part of the Hilbert space contributes to the problem at hand.
We are going to assume that all relevant configurations contain
only one heavy particle; in other words, in the classical limit
no new black holes are created in this $S$-matrix evolution.
Let us also exclude decay of the heavy particle,
\ie black-hole evaporation, from current consideration.
In other words, we assume that the spectral density\footnote{In the
   coherent-spin approach to the classical limit~\cite{Aoude:2021oqj},
   the ${\rm SU}(2)$ spinors $\beta_b$ that saturate the little-group indices of the amplitude
   determine the resulting classical angular momentum of the compact object.
   So one could trade the $s_2$-dependence of the spectral density in \eqn{Completeness}
   for the perhaps more appropriate dependence on $\hbar\|\beta\|^2 = 2 \sqrt{-S_\text{cl}^2}$
   and use the coherent-spin final-state integration.
   All formulae in this section may be adjusted accordingly, starting with \eqn{eq:ProbCons} which becomes
\be
P^\text{(c)}_{\gamma \to \tilde{\gamma}}
 = \int_{p_2}\!\int\!\frac{d^4\beta}{\pi^2} \bra{\rin} S^\dagger
   \ket{p_2,\beta;\tilde{\gamma}^{\tilde{h}}}
   \bra{p_2,\beta;\tilde{\gamma}^{\tilde{h}}} S \ket{\rin} . \nn
\ee
Note that as long as the integration over the ${\rm SU}(2)$ spinors $\beta_b$
appears in the final-state summation,
one may regard and use it as a shorthand for the bulkier spin sum.
} of the heavy-particle states has a non-trivial continuous part only for $M_2>M_1$
(alongside the delta-function responsible for case $\text{(c)}$):
\be
\label{Completeness}
1^\text{(nc)}
 = \sum_{X_\text{rad}} \sum_{s_2} \sum_{\{b\}}
   \int_{M_1^2}^\infty\!dM_2^2 \rho_{s_2}(M_2^2) \int_{p_2}\!
	\ket{p_2,s_2,\{b\}; X_\text{rad}} \bra{p_2,s_2,\{b\} ; X_\text{rad} } .
\ee

The above ``completeness'' relation should normally also include a sum
over possible emitted radiation
\be
\ket{X_\text{rad}} \bra{X_\text{rad}}
 = \sum_{n=0}^\infty \sum_{h_1, \cdots, h_n} \int_{k_1,\cdots,k_n}
   \ket{k_1^{h_1}; \cdots; k_n^{h_n}} \bra{k_1^{h_1}; \cdots; k_n^{h_n}} .
\ee
However, we choose to make another assumption that
all the outgoing radiation belongs coherently to the wave $\tilde{\gamma}$,
and there is no extra scattered photons/gravitons.
In other words, the final state is given by
$\ket{p_2,\beta;\tilde{\gamma}^{\tilde{h}}}$ and not
$\ket{p_2,\beta;\tilde{\gamma}^{\tilde{h}};k_1^{h_1};k_2^{h_2};\cdots}$,
which was also assumed for the mass-preserving case~\eqref{eq:ProbCons}.
This assumption relies on the expectation that radiated quanta
are not classically significant unless they belong to a classical wave
modeled by a coherent state, see \eg \cite{Cristofoli:2021jas}.
Therefore, remembering the meaning of the incoming state,
we can write the absorption probability as
\begin{align}
\label{eq:ProbAbs}
P^\text{(nc)}_{\gamma \to \tilde{\gamma}}
 = & \int_{p_1,p_1'}\!\!\psi_\xi^*(p_1) \psi_\xi(p_1')
   e^{ib\cdot (p_1'-p_1)} \\ \times
   \sum_{s_2} & \int_{M_1^2}^\infty\!dM_2^2 \rho_{s_2}(M_2^2)
   \int_{p_2}\!\sum_{\{b\}}
   \bra{p_1;\gamma^h} S^\dagger \ket{p_2,s_2,\{b\};\tilde{\gamma}^{\tilde{h}}}
   \bra{p_2,s_2,\{b\};\tilde{\gamma}^{\tilde{h}}} S \ket{p_1';\gamma^h} . \nn
\end{align}
The building block
$\bra{p_2,s_2,\{b\};\tilde{\gamma}^{\tilde{h}}} S \ket{p_1;\gamma^h}$
involves a transition of a scalar heavy state into a possibly spinning one
along with the incoming and outgoing massless coherent states.
Since the latter states contain an infinite number of photons/gravitons,
the matrix elements of $S=1+iT$ should be expanded in perturbation theory.

\subsection{$T$-matrix perturbative expansion}

The massless coherent states (plane or spherical) are sensitive to
all orders in perturbation theory,
and their matrix elements are non-trivial~\cite{Cristofoli:2021vyo}.
However, we can expand operators
in terms of annihilation and creation operators, plane or spherical.
We are going to perform the $T$-matrix expansion
in the following way:\footnote{We thank Donal O'Connell
for valuable discussions on the expansion~\eqref{eq:TmatrixNotation}.}
\begin{align}
\label{eq:TmatrixNotation}
 T = \sum_{m,n=0}^\infty
    \Big(T_{(m|n)}^\text{(c)} + T_{(m|n)}^\text{(nc)} \Big)
 & = T_{(0|1)}^\text{(nc)} + T_{(1|0)}^\text{(nc)} \\
+\:T_{(1|1)}^\text{(c)} + T_{(0|2)}^\text{(c)} + T_{(2|0)}^\text{(c)} &
 + T_{(1|1)}^\text{(nc)} + T_{(0|2)}^\text{(nc)} + T_{(2|0)}^\text{(nc)}
 + \cdots , \nn
\end{align}
where the superscripts $\text{(c)}$ and  $\text{(nc)}$ represent mass-preserving and mass-changing elements, respectively,
while the subscript $(m|n)$ corresponds to $n$ incoming
and $m$ outgoing photons/gravitons,
and each $T$-matrix element will generate an $(m+n+2)$-point amplitude. 
In the first line of \eqn{eq:TmatrixNotation}, we have isolated
the leading non-conservative effects due to absorption,
$T_{(0|1)}^\text{(nc)}$, and emission, $T_{(1|0)}^\text{(nc)}$.
Both terms are mass-changing three-point amplitudes and non-zero
even on real kinematics,
while the mass-preserving counterparts vanish,
$T_{(1|0)}^\text{(c)} = T_{(0|1)}^\text{(c)} = 0$.\footnote{See
\cite{Cristofoli:2022phh} for a discussion of large gauge effects,
where such amplitudes do contribute.
}
In this paper, we have been studying the leading-order in absorption term $T_{(0|1)}^\text{(nc)}$,
but the above expansion allows to also systematically understand higher orders.

In the second line, we have four-point terms that lead to the usual conservative Compton amplitude $T_{(1|1)}^\text{(c)}$
and its non-conservative counterpart $T_{(1|1)}^\text{(nc)}$.
The former has been vastly studied recently,
but the latter has been unexplored to the best of our knowledge.
Furthermore, we have double-emission $(2|0)$ and double-absorption $(0|2)$
both on the conservative and non-conservative sides.
Together with the non-conservative Compton,
double-absorption would give the naive next-to-leading order (NLO) terms
to our leading-order analysis.

The $T$-matrix elements can be written in terms of scattering amplitudes:
\beal
T_{(m|n)} & =\!\!\sum_{2s_1,2s_2=0}^\infty \int_{p_1,p_2}\!
	\sum_{\substack{h_1,\dots,h_n \\ \tilde{h}_1,\dots,\tilde{h}_m}}\!\!
	\int_{\,\substack{k_1,\dots,k_n \\ \tilde{k}_1,\dots,\tilde{k}_m}}\!
	\hat{\delta}^4\big( p_1\!+\!{\textstyle \sum_i} k_i
                     - p_2\!-\!{\textstyle \sum_i} \tilde{k}_i
                 \big) \\ & \qquad \times
	{\cal A}_{\{b\}}{}^{\{a\}}(p_2,s_2;\tilde{k}_1,\tilde{h}_1; \dots;
	\tilde{k}_m,\tilde{h}_m|p_1,s_1;k_1,h_1; \dots; k_n,h_n) \\ & \qquad\times\!
	\Big[ a^{\dagger \{b\}}(p_2,s_2)
	      a_{\tilde{h}_1}^\dagger(\tilde{k}_1) \cdots
	      a_{\tilde{h}_m}^\dagger(\tilde{k}_m) \Big]\!
	\Big[ a_{\{a\}}(p_1,s_1) a_{h_1}(k_1) \cdots a_{h_n}(k_n) \Big] ,
\label{eq:TmatrixExpansion}
\eeal
where for brevity we have left the summation
over the symmetrized massive little-group indices
$a_1,\dots,a_{2s_1}$ and $b_1,\dots,b_{2s_2}$ implicit.
The integration measure over $p_2$ contains either
$\delta^+(p_2^2-M_1^2)$ or $\delta^+(p_2^2-M_2^2)$,
depending on the conservative or non-conservative sector considered.
The corresponding two sets of $T$-matrix elements span the space of
one heavy particles and all possible photon/graviton radiation
being emitted and absorbed.
Of course, this is not the whole $T$-matrix space,
since we could have more heavy particles
and a mixed combination of photons and gravitons,
but here we restrict to only one messenger particle.

For simplicity, we have used plane-wave massless
creation/annihilation operators, which return the waveshape $\gamma(k)$
when applied to plane-wave coherent states:
\be
a_h(k) \ket{\gamma^{h'}} = \gamma(k) \delta_h^{h'} \ket{\gamma^h} ,
\ee
see \eg \cite{Cristofoli:2021vyo}.
Aiming for the leading-order absorption effects,
we can evaluate the contribution of the $T_{(0|1)}^\text{(nc)}$
matrix element to the mass-changing probability as
\beal
\label{eq:InXAmp}
\braket{p_2,s_2,&\{b\};\tilde{\gamma}^{\tilde{h}}| S |p_1;\gamma^h}
 \simeq i\braket{p_2,s_2,\{b\};\tilde{\gamma}^{\tilde{h}}|
   T_{(0,1)}^\text{(nc)} |p_1;\gamma^h} \\ &
 = i\!\int_k\!\hat{\delta}^4(p_1+k-p_2)
   {\cal A}_{\{b\}}(p_2,s_2|p_1;k,h')
   \braket{\tilde{\gamma}^{\tilde{h}}|a_{h'}(k)|\gamma^h} \\ &
 = i\delta^h_{\tilde{h}} \braket{\tilde{\gamma}^h|\gamma^h}\!
   \int_k\!\gamma(k)\hat{\delta}^4(p_1+k-p_2)
   {\cal A}_{\{b\}}(p_2,s_2|p_1;k,h) .
\eeal

\subsection{Partial-wave coherent states}
\label{sec:SphCoherent}

We are interested in the scattering of a partial wave with a black hole,
with the wave modeled by a covariant spherical coherent state.
Such states are defined as eigenstates of the spherical annihilation operators:
\be
a_{j,m,h}(\omega) \ket{\gamma^{h'}}
 = \gamma_{j,m}(\omega) \delta_h^{h'} \ket{\gamma^h} , \qquad
\ket{\gamma^h} = {\cal N}_\gamma
	\exp\!\bigg[ \sum_{j,m}\!\int_0^\infty\!\hat{d}\omega\,\gamma_{j,m}(\omega)
                a^\dagger_{j,m,h}(\omega) \bigg] \ket{0} .
\ee
Setting $\braket{\gamma^h|\gamma^h}=1$ gives the normalization prefactor as
\be
{\cal N}_\gamma = \exp\!\bigg[{-}\frac{1}{2} \sum_{j,m}\!
   \int_0^\infty\!\hat{d}\omega\,|\gamma_{j,m}(\omega)|^2 \bigg] .
\ee
The waveshape $\gamma_{j,m}(\omega)$ of these coherent states
describes the contribution of each $(j,m)$ component to the total wave,
and we expect that in the classical limit $\gamma_{j,m}(\omega)$ is peaked
at the frequency $\omega_\text{cl}$.
We can simplify the problem further by studying the incoming wave
$\ket{\gamma_{j,m}^h}$ with just a particular $(j,m)$ component,
in which case the spherical waveshape reduces to
$\gamma_{j',m'}(\omega) = \delta^j_{j'} \delta^m_{m'} \gamma(\omega)$,
such that
\be
a_{j,m,h}(\omega) \ket{\gamma_{j',m'}^{h'}} = \gamma(\omega)
\delta_j^{j'} \delta_m^{m'} \delta_h^{h'} \ket{\gamma_{j,m}^h} .
\ee

Coming back to the initial state $\ket{\rin}$ given in \eqn{eq:InState},
which describes a scalar black hole and a partial wave
as a wavepacket superposition of $\ket{p_1;\gamma_{j,m}^h}$.
The $S$-matrix determines the probability amplitude of its evolution
into a final massive state~$X$ and another partial wave
$\ket{\tilde{\gamma}^{\tilde{h}}}$
with perhaps more than one $(\tilde{\jmath},\tilde{m})$ components.
Let us write the leading absorption term $T_{(0|1)}^\text{(nc)}$
to such a process,
by switching the states on the left-hand side of \eqn{eq:InXAmp}
from plane to spherical waves:
\be
\label{eq:InXAmpSph0}
\braket{p_2,s_2,\{b\};
   \tilde{\gamma}^{\tilde{h}}|S|p_1;\gamma_{j,m}^h}
\simeq i\!\!\int_k\!\hat{\delta}^4(p_1+k-p_2)
   {\cal A}_{\{b\}}(p_2,s_2|p_1;k,h')
   \braket{ \tilde{\gamma}^{\tilde{h}}|a_{h'}(k)|\gamma_{j,m}^h } .
\ee
The main difference is that to evaluate the matrix element of a plane-wave annihilation operator between two spherical coherent states,
we need to summon the decomposition of the plane-wave operator
into partial waves:
\be
a_h(k) = 4\pi \sum_{j=|h|}^\infty \sum_{m=-j}^j
   \int_0^\infty\!\!\frac{\hat{d}\omega}{\!\sqrt{2\omega}}
   \hat{\delta}(k \cdot u_1-\omega)\,{}_{-h}Y_{j,m}(k;u_1) a_{j,m,h}(\omega) ,
\ee
and hence
\be
a_{h'}(k) \ket{\gamma_{j,m}^h}
 = \frac{4\pi \delta_{h'}^h}{\!\sqrt{2k\cdot u_1}}
   \gamma_{j,m}(k\!\cdot\!u_1)\,{}_{-h}Y_{j,m}(k;u_1) \ket{\gamma_{j,m}^h} .
\ee
Therefore, we compute the leading mass-changing matrix element as
\begin{align}
\label{eq:InXAmpSph}
\braket{ p_2,s_2,\{b\};\tilde{\gamma}^{\tilde{h}}|S|p_1;\gamma_{j,m}^h }
\simeq 4\pi i \braket{ \tilde{\gamma}^{\tilde{h}}|\gamma_{j,m}^h }\!\!
   \int_0^\infty\!\!\frac{\hat{d}\omega}{\!\sqrt{2\omega}}
   \gamma_{j,m}(\omega) & \\ \times\!
   \int_k\!\hat{\delta}(k \cdot u_1-\omega)\,{}_{-h}Y_{j,m}(k;u_1)
   \hat{\delta}^4(p_1+k-p_2) & {\cal A}_{\{b\}}(p_2,s_2|p_1;k,h) . \nn
\end{align}

The leading contribution to the absorption probability~\eqref{eq:ProbAbs}
is then given by
\begin{align}
\label{eq:ProbAbsSph}
P^\text{(nc)}_{\gamma \to \tilde{\gamma}} & \simeq 8\pi^2
   \big| \braket{ \tilde{\gamma}^{\tilde{h}}|\gamma_{j,m}^h } \big|^2
   \sum_{s_2}\!\int_{M_1^2}^\infty\!dM_2^2 \rho_{s_2}(M_2^2)\!
   \int_{p_1,p_1',k,k'\!,p_2}\!\!\psi_\xi^*(p_1) \psi_\xi(p_1')
   e^{ib\cdot (p_1'-p_1)} \\ & \times\!
   \int_0^\infty\!
   \frac{\hat{d}\omega \hat{d}\omega'}{\sqrt{\omega \omega'}}
   \gamma^*(\omega) \gamma(\omega')
   \hat{\delta}(k \cdot u_1\!-\!\omega) \hat{\delta}(k'\!\cdot u_1\!-\!\omega')
   \hat{\delta}^4(p_1+k-p_2)
   \hat{\delta}^4(p_1'\!+k'\!-p_2) \nn \\ & \qquad \quad~\;\:\times
   {}_{-h}Y_{j,m}^*(k;u_1)\,{}_{-h}Y_{j,m}(k';u_1)\,
   {\cal A}^{*\{b\}}(p_2,s_2|p_1;k,h)\,
   {\cal A}_{\{b\}}(p_2,s_2|p_1';k',h) . \nn
\end{align}
Note that apart from the overlap between the two spherical coherent states
and the impact-parameter exponent, we have landed exactly on
the single-quantum absorption cross-section
given in \eqns{eq:SphHelProb2}{eq:SphHelProb2A} ---
with the $(j,m)$ waveshape $\gamma(\omega)$ as the single-particle
energy wavefunction.
In other words, we observe that 
the waveshape $\gamma(\omega)$ acts as a one-dimensional wavefunction,
which smears the energy spectrum but is peaked at the classical frequency
$\omega_\text{cl}$.
This observation was also made in~\cite{Endlich:2016jgc},
where single quanta and coherent states gave the same results.

Let us discuss the seeming discrepancies between
the leading-order cross-section formulae~\eqref{eq:SphHelProb2}
and \eqref{eq:ProbAbsSph}.
For a spherical wave defined in the rest-frame
of (the classical momentum of) the compact body and centered at it,
the impact parameter should of course be set to zero.
Moreover, \eqns{eq:SphHelProb2}{eq:SphHelProb2A}
were written for an inclusive probability,
let us rename it to
$P_{(0|1)}^\text{(nc)} := P_\text{inc}^\text{LO}(\omega_\text{cl},j,m,h)$,
whereas retaining the dependence on the outgoing waveshape
in \eqn{eq:ProbAbsSph} is actually an enhancement
of the single-quantum formula:
\be
P^\text{(nc)}_{\gamma\to\tilde{\gamma}}
 = \big| \braket{\tilde{\gamma}^{\tilde{h}}|\gamma_{j,m}^h} \big|^2
   P_{(0|1)}^\text{(nc)} + \dots ,
\ee
where the dots denote the higher orders to be briefly discussed below.
In the limit where the outgoing classical wave changes very little,
the above prefactor may furthermore disappear,
$\braket{ \tilde{\gamma}^{\tilde{h}}|\gamma_{j,m}^h } \approx 1$.

\subsection{Higher-order diagrammatics}
\label{sec:Diagrammatics}

In this section, we use diagrams to help us understand all the effects
relevant for BH-wave interactions.
Having a diagrammatic realization of the expressions from the previous sections will guide us for the NLO corrections.
However, this diagrammatic approach is general enough to be also applicable
to any order in perturbation theory,
as well as such processes as emitted radiation and superradiance.

\begin{figure}[!htb]
\centering
\vspace{-5pt}
\includegraphics[width=15cm]{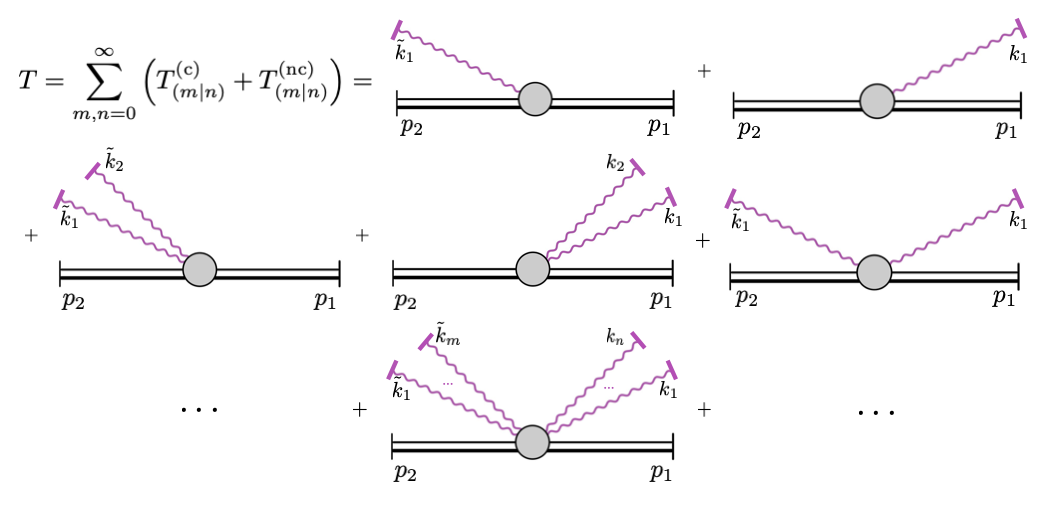}
\vspace{-25pt}
\caption{$T$-matrix operator expansion}
\label{fig:Tmatrix_diags}
\end{figure}

Let us take a brief moment to explain the diagrammatic expansion
of $T$-matrix in \fig{fig:Tmatrix_diags},
which represents \eqn{eq:TmatrixNotation}.
The operator nature of this diagram is represented by the ``vertical line'' after the wavy graviton line, and the double lines,
which will ``act'' on a ket quantum state,
\eg the massless coherent state $\ket{\gamma^h}$
or the black-hole $\ket{p_1}$.
In this diagram, we then have
\begin{itemize}[leftmargin=21pt,parsep=0pt]
\item
$n$ incoming graviton annihilation operators shown by wavy lines and labeled by $\{k_1,\cdots,k_n\}$;
\item
$m$ outgoing graviton creation operators shown by wavy lines and labeled by $\{\tilde{k}_1,\cdots,\tilde{k}_m\}$;
\item
incoming and outgoing double line, labeled by $p_1$ and $p_2$.
The two lines of different thickness inside of the double line represents the fact that this diagram contains mass-preserving and mass-changing transitions.
\item
vertical lines at the end of graviton/BH lines represent the operator nature of these diagrams. For instance, the double-line part of the operator will act on $|p_1\rangle$, while the wavy line will act on the coherent state $|\gamma^h\rangle$.
\item
Evaluating this operator with outgoing states on the left and incoming states on the right will result in scattering amplitudes, waveshapes, and coherent-state overlap. Due to the operator-action convention, time flows from right to left in the resulting amplitude.
\end{itemize} 

\begin{figure}[t]
\centering
\vspace{-5pt}
\includegraphics[width=14cm]{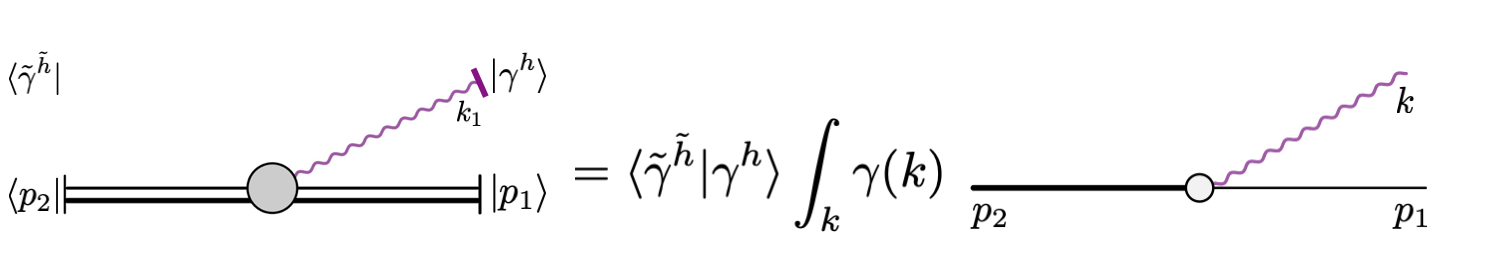}
\vspace{-15pt}
\caption{$T_{(0|1)}^\text{(nc)}$-matrix operator acting on the quantum states. Time flows right to left.
}
\label{fig:SingleAbsorptionTmatrixOperator}
\end{figure}

Let us now apply these diagrams to the evaluation
of the leading-order contribution to absorption given in \eqn{eq:InXAmp}.
We take the first term $T_{(0|1)}^\text{(nc)}$ on the right-hand side
of \fig{fig:Tmatrix_diags} and take its matrix element
$\braket{p_2,s_2,\{b\};\tilde{\gamma}^{\tilde{h}}|
  T_{(0,1)}^\text{(nc)} |p_1;\gamma^h}$.
The result is the overlap between the coherent states, a scattering amplitude, and the waveshape $\gamma(k)$, represented in \fig{fig:SingleAbsorptionTmatrixOperator}. Note that the integrated scattering amplitude is a single-graviton amplitude smeared by the waveshape.

\begin{figure}[b]
\centering
\vspace{-5pt}
\includegraphics[width=11cm]{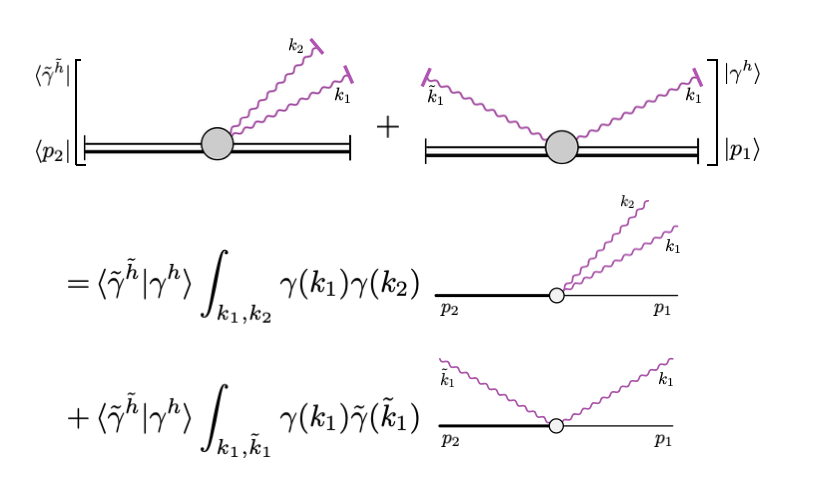}
\vspace{-15pt}
\caption{Next-to-leading order contributions to mass-changing absorption effects}
\label{fig:AbsorptionTmatrixOperatorNLO}
\end{figure}

Similarly, \fig{fig:AbsorptionTmatrixOperatorNLO} shows
how this diagrammatic technique applies to
the NLO non-conservative contributions.
They contains double absorption and the mass-changing Compton amplitude,
which both involve two photons/gravitons,
now integrated with two waveshapes coming from the coherent states.

\subsection{PM absorption analysis}
\label{sec:HigherPM}

In the previous section, we have explained how
to include higher orders in multiplicity into the BH-wave interaction modeling
by expanding the $T$-matrix.
The PM expansion, however, enters into the mass-changing amplitudes
in a rather intricate way.
Indeed, as we have seen from \eqn{eq:MatchingLO},
even the three-point absorptive amplitudes must behave ${\cal O}(G^{s_2+1})$.
Let us now explore the mass-changing $(m+n+2)$-point amplitude
${\cal A}_{\{b\}}{}^{\{a\}}(p_2,s_2;\tilde{k}_1,\tilde{h}_1; \dots;\tilde{k}_m,\tilde{h}_m|p_1,s_1;k_1,h_1; \dots; k_n,h_n)$
in \eqn{eq:TmatrixExpansion}.
For brevity, we compress the notation to
${\cal A}^{(s_2|s_1)}_{\text{abs}(m|n)}$,
emphasizing its distinction from the mass-conserving counterparts
${\cal A}^{(s_2|s_1)}_{(m|n)}$.
In particular, at three points we have
\be
{\cal A}^{(s_2,0)}_{\text{abs}(0|1)} \propto G^{s_2+1}, \qquad \quad
{\cal A}^{(s)}_{3,\text{min}} \propto \sqrt{G} ,
\ee
where the second one is the usual three-point same-mass
amplitude~\cite{Arkani-Hamed:2017jhn}
of the minimal form~\eqref{eq:AHHStripped}, which are known to
correspond to Kerr BHs at 1PM \cite{Guevara:2018wpp,Chung:2018kqs}.

To obtain higher multiplicities, we can now naively multiply the powers of the Newton constant of these three-point amplitudes,
assuming that they scale uniformly in $G$,
and any subleading orders at three points
should come from higher loop orders.\footnote{See \cite{Saketh:2023bul} for loop corrections to Love numbers in the worldline EFT framework. For quantum corrections to Love numbers  due to emission see \cite{Kim:2020dif},
which we also ignore in the above analysis.
}
At four points, we have two incoming gravitons and a mass-changing heavy particle.
We then have three types of contributions: a contact four-point term,
two successive three-point absorptions,
and one absorption together with one minimal-coupling amplitude.
These terms be written respectively as
\be
\label{eq:AbsAmplitudeNLO}
{\cal C}^{(s_2,0)}_{{\rm abs}(0|2)}\,+\,
\underbrace{{\cal A}^{(s_2,0)}_{\text{abs}(0|2)+0}}_{\propto\,G^{2s_2+2}}\,+\,
\underbrace{{\cal A}^{(s_2,0)}_{\text{abs}(0|1)+1}}_{\propto\,G^{s_2+3/2}}
 =: {\cal A}^{(s_2,0)}_{\text{abs}(0|2)} ,
\ee
where the subscript notation $(0|r)+n-r$ means that we have $n$ gravitons,
$r$ out which couple via an absorptive three-point amplitude
and $(n-r)$ via the mass-preserving minimal coupling.
More generally, for $n$-graviton absorption we thus have
\be\!\!\!
{\cal A}^{(s_2,0)}_{\text{abs}(0|n)}\!
 = \sum_{r=1}^{n} {\cal A}^{(s_2,0)}_{{\rm abs}(0|r)+n-r}
 +\,{\cal C}^{(s_2,0)}_{\text{abs}(0|n)} , \qquad
{\cal A}^{(s_2,0)}_{\text{abs}(0|r)+n-r}\!
 \propto G^{r(s_2+1)+(n-r)/2} .
\ee

In \sec{sec:Matching}, we have seen that, on the GR side, the PM expansion
of the near-zone response function~\eqref{eq:ResponseFunction} suggests that
the leading-order absorption cross-section scales as $G^{2j+2}$,
whereas the NLO does as $G^{2j+4}$.\footnote{Tail effects may modify the NLO to ${\cal O}(G^{2j+2})$
\cite{Saketh:2022xjb,Saketh:2023bul},
but we expect them to arise from loops.
}
Now from squaring the amplitudes~\eqref{eq:AbsAmplitudeNLO},
we see that we obtain terms that scale as
$G^{2j+3}$, $G^{3j+7/2}$ and $G^{4j+4}$
for $s_2=j$ (as follows from spin conservation seen in \eqn{eq:SphHelAmplitudeCovMin}).
Therefore, it is not possible to obtain the NLO $G^{2j+4}$
expected on the GR side from the tree-level counting on the EFT side,
unless the contact term is artificially introduced to account for this counting.
However, a more natural way to obtain the expected behavior in $G$ is
from the amplitude with three incoming gravitons, which is expanded as
\be
{\cal A}^{(s_2,0)}_{{\rm abs}(0|3)} 
 = \underbrace{{\cal A}^{(s_2,0)}_{\text{abs}(0|1)+2}}_{\propto\,G^{s_2+2}}
 + \underbrace{{\cal A}^{(s_2,0)}_{\text{abs}(0|2)+1}}_{\propto\,G^{2s_2+5/2}}
 + \underbrace{{\cal A}^{(s_2,0)}_{\text{abs}(0|3)+0}}_{\propto\,G^{3s_2+3}}
 + {\cal C}^{(s_2,0)}_{{\rm abs}(0|3)} .
\ee
Indeed, we see that the first contribution squared induces
the desired NLO $G^{2j+4}$ correction to the absorption cross-section.

\section{Summary and discussion}
\label{sec:Outro}

In this work, we have initiated the exploration of classical absorption effects
for compact bodies using quantum scattering amplitudes.
Central to this program are the mass-changing three-point scattering amplitudes \cite{Arkani-Hamed:2017jhn,Conde:2016vxs}
that entail new degrees of freedom modeling non-conservative effects,
which may change the mass and spin of the heavy particle
(representing the compact object) due to the incoming wave.

We have made use of these amplitudes and their connection
to covariantized spin-weighted spherical harmonics to describe
leading gravitational absorption effects from a macroscopic/EFT point of view.
Since this is an effective description,
matching to the underlying theory was required to obtain the values
of the EFT coupling coefficients.
We have chosen to match at the cross-section level
to the GR calculation dating back to Starobinsky, Churilov~\cite{Starobinsky:1973aij,Starobinskil:1974nkd} and Page~\cite{Page:1976df,Page:1976ki}.
Although we have performed a leading-order match,
this probability-based formalism can accommodate higher orders
in the PM expansion and incoming spinning BHs and neutron stars as well.
For the latter case, absorption effects were considered
via tidal heating~\cite{1992ApJ...397..570M,Lai:1993di},
and it would be interesting to understand how the effective couplings $g_{r,s_1,s_2}$ deviate from the BH values. We leave this for future work. 

Having made sense of the effective couplings,
we have explored how the used single-quantum framework fits into
a more general and consistent description of classical waves
using massless coherent states.
In particular, we were able to connect the frequency wavefunction
used in the former with the coherent-state waveshape,
\ie the eigenvalue of the annihilation operator.
An interesting feature of this analysis is the diagrammatic approach
for expanding the $T$-matrix and systematically introducing
higher-order terms in the coherent cross-section.
Crucial to this analysis was the separation of the probabilities
into conservative and absorptive,
which is motivated by the intrinsically distinct nature
of the quantum amplitudes building blocks. 
Although the classical limit sends $M_2 \to M_1$,
the form of the resulting cross-section follows from the amplitudes
constructed on $M_2 \neq M_1$ kinematics,
which are qualitatively different from their same-mass counterparts, since they belong to distinct Hilbert spaces.

The natural next step is to include spin effects for the initial black hole with the end goal of modeling a Kerr BH absorption cross-section purely from on-shell amplitudes.
According to the microscopic calculation from the GR side,
such leading-order non-spinning effects come at ${\cal O}(G^3)$
at the cross-section level, suggesting that the effective coupling
in the amplitude should start at ${\cal O}(G^{3/2})$.
From the EFT side, in this more general case of $s_1 \neq 0$,
we have observed the proliferation of possible effective couplings
in the mass-changing three-point amplitude~\eqref{eq:ThreePointGeneral}, making the matching a harder task.
However, the proposed definition of the mass-changing minimal amplitudes~\eqref{eq:ThreePointPlusAmpMin} might streamline the calculation
and perhaps even correspond to the Kerr BH in the same way
as the same-mass ``minimal coupling''~\cite{Arkani-Hamed:2017jhn}
of the form~\eqref{eq:AHHStripped} are known to~\cite{Guevara:2018wpp,Chung:2018kqs}.

Another direction that we have not explored is the study of observables
from amplitudes, in particular using
the KMOC formalism~\cite{Kosower:2018adc,Maybee:2019jus,delaCruz:2020bbn,Cristofoli:2021vyo,Aoude:2021oqj,Cristofoli:2021jas}. 
With the obtained absorption effective coefficients,
many interesting local and global observables could be already be explored at leading or higher PM orders using the presented formalism.
Perhaps the most interesting ones are the change in mass and spin
induced by absorption, where one could naturally use such quantum operators
as  $\mathbb{P}^2 = \mathbb{P}^\mu \mathbb{P}_\mu $ to obtain $\Delta M^2$
and $\mathbb{S}^2 =\mathbb{S}^\mu\mathbb{S}_\mu$ to obtain $\Delta S^2$.
Moreover, one could imagine probing the change in the area of the BH
due to absorptive effects. In classical GR, the area is defined as
\be
A_\text{H} := 8\pi (G M)^2\bigg[1+ \sqrt{1-\chi^2}\bigg], \qquad
\chi = \frac{\mathfrak{a}}{G M},
\ee
and $\mathfrak{a} = \sqrt{-S^2}/M$ is the Kerr ring radius.
To obtain the change in this quantity from amplitudes, 
one would like to define a QFT operator for the area and try to compute $\Delta A_\text{H}$ in a scattering process. 
For that, one could substitute $(S^2,M^2) \to (\mathbb{S}^2,\mathbb{P}^2)$,
which imples the following proposal for the area operator:
\be
\mathbb{A}_\text{H} = 8\pi \left[ G^2\,\mathbb{P}^2
 + \sqrt{(G^2\,\mathbb{P}^2)^2 + G^2\,\mathbb{S}^2 }\right] ,
\ee
which mixes PM orders.
The simplicity of this proposal also comes from the fact that 
the two operators commute $[\mathbb{S}^2,\mathbb{P}^2]=0$.
The mixing between orders in the expansion brings an 
interesting interplay between the $\mathbb{S}^2$ and the $\mathbb{P}^2$ calculations.
We leave the exploration of such an operator for future work.

We hope that this work may open these and other avenues to include absorption effects in the on-shell amplitude approach to gravitational waves.
In particular, the work~\cite{Bautista:2022wjf} on matching
Teukolsky-equation solutions to the gravitational Compton scattering amplitudes suggests that absorption effects could be included into them
in relation to horizon effects.
It is tempting to consider these effects from a purely on-shell perspective,
as the four-point amplitudes are likely to be related
to the leading-order absorption cross-section by dispersion relations.

Another direction is to explore in more detail
the role of the spectral density function that we were forced to introduce
in our formalism.
For instance, it would be interesting to see if it appears in a similar way
in the context of the Heavy Particle Effective Theory~\cite{Damgaard:2019lfh,Aoude:2020onz}, which streamlines the classical limit.
We leave this also for future work.

\section*{Acknowledgements}

We are grateful to Fabian Bautista, Kays Haddad, Andreas Helset, Yu-tin Huang, Jung-Wook Kim, Nathan Moynihan, Donal O'Connell and M.V.S. Saketh for valuable conversations, especially to Andreas, Donal and Jung-Wook
for the comments of an early draft of this paper.
RA's research was supported by the F.R.S.-FNRS project no. 40005600 and the FSR Program of UCLouvain.

\appendix

\section{Spherical harmonics and spinors}
\label{app:SphHarmCov}

Here we discuss the spinorial construction for
the spin-weighted spherical harmonics.

\paragraph{Spherical harmonics in 3d.}
The original construction due to Newman and Penrose~\cite{Newman:1966ub}
may be neatly formulated (see \eg \cite{Torres:2007})
in terms of ${\rm SU}(2)$ spinors on the sphere
$S^2 = \{ \hat{\bm{k}} = (\cos\varphi \sin\theta,\sin\varphi \sin\theta,\cos\theta) \} \subset \mathbb{R}^3$:
\be\!\!
   \kappa_+^a\!
 = \begin{pmatrix}
      e^{-\frac{i\varphi}{2}}\!\cos\frac{\theta}{2} \\
      e^{\frac{i\varphi}{2}}\!\sin\frac{\theta}{2}
   \end{pmatrix}\!,~~
   \kappa_-^a\!
 = \begin{pmatrix}\!
     -e^{-\frac{i\varphi}{2}}\!\sin\frac{\theta}{2} \\
      e^{\frac{i\varphi}{2}}\!\cos\frac{\theta}{2}
   \end{pmatrix} ~~\Rightarrow~~
   \Bigg\{
   \begin{aligned}
    & \hat{\bm{k}} \cdot \bm{\sigma}^a{}_b\,\kappa_\pm^b = \pm \kappa_\pm^a , \\
    & \hat{k}^i\!= -\frac{1}{2} \sigma^{i,a}{}_b
         (\kappa_+^a \kappa_{-b} + \kappa_-^a \kappa_{+b}) ,\!
   \end{aligned}\!
\label{Spinors3d}
\ee
where $\bm{\sigma}^a{}_b$ is the concatenation
of the three standard Pauli matrices.
We then define
\be
{}_h\tilde{Y}_{j,m}(\hat{\bm{k}})
:= \lefteqn{ \overbrace{\phantom{\kappa_+^{(1}\!\cdots\kappa_+^1}\!
             }^{j-m} }
   \lefteqn{ \underbrace{\phantom{\kappa_+^{(1}\!\cdots\kappa_+^1
                                  \kappa_+^2\!\cdots\zeta_+^2}
             }_{j+h} }
   \kappa_+^{(1}\!\cdots\kappa_+^1
   \overbrace{ \kappa_+^2\!\cdots\kappa_+^2
               \underbrace{\kappa_-^2\!\cdots\kappa_-^{2)}}_{j-h} }^{j+m} .\label{SphHarmSpinors3d}
\ee
Up to normalization, these functions are directly related
to the conventional angle-dependent harmonics \cite{Goldberg:1966uu}
via the spinor parametrization~\eqref{Spinors3d}:
\begin{align}
\label{SphHarmGoldberg}
 & {}_hY_{j,m}(\theta,\varphi) := (-1)^m
   {\textstyle \sqrt{\frac{(2j+1)(j+m)!(j-m)!}{4\pi(j+h)!(j-h)!}}}
   \big(\!\sin\tfrac{\theta}{2}\big)^{2j}\!\!
   \sum_{r={\rm max}(0,m-h)}^{{\rm min}(j+m,j-h)}\!\!(-1)^{j-h-r}
   {\textstyle {j-h \choose r}} \\ &
   \times {\textstyle {j+h \choose r+h-m}} 
   \big(\!\tan\tfrac{\theta}{2}\big)^{m-h-2r} e^{im\varphi}
 = (-1)^m (2j)!
   \textstyle{\sqrt{\frac{2j+1}{4\pi (j+m)!(j-m)! (j+h)!(j-h)!}}}\;
   {}_h\tilde{Y}_{j,m}(\hat{\bm{k}}) . \nn
\end{align}
The latter functions obey the standard orthonormality property on $S^2$,
\be
\int\!d\Omega_{\hat{\bm{k}}}\;
   {}_hY^*_{j',m'}(\hat{\bm{k}})\;{}_hY_{j,m}(\hat{\bm{k}})
 = \delta_j^{j'} \delta_m^{m'} .
\label{SphHarmNorm}
\ee
Note that the usual conventions~\eqref{SphHarmGoldberg} fix
the (functional) ${\rm U}(1)$ freedom 
\be
\kappa_\pm \to e^{\pm i\phi(\bm{\hat{k}})/2} \kappa_\pm
\qquad \Rightarrow \qquad
{}_hY_{j,m}(\hat{\bm{k}})
   \to e^{ih\phi(\hat{\bm{k}})} {}_hY_{j,m}(\hat{\bm{k}}) ,
\label{SphHarmRotationU1}
\ee
which leaves the directional vector $\bm{\hat{k}}$ invariant
and does not affect any important properties of the spherical harmonics.

\paragraph{Spinors in 4d.}
In Minkowski space, spinors carry ${\rm SL}(2,\mathbb{C})$
Weyl indices $\alpha$ or~$\dot{\alpha}$
of negative or positive chirality, respectively.
In the spinor-helicity formalism,
spinors are denoted by bras and kets, written
with angle or square brackets depending on their chirality.
The spinors corresponding to massless vectors obey~\cite{Berends:1981rb,DeCausmaecker:1981jtq,Gunion:1985vca,Kleiss:1985yh,Xu:1986xb,Gastmans:1990xh}
\be
   \ket{k}_{\alpha} [k|_{\dot{\alpha}}
    = k_{\alpha \dot{\alpha}}
   := k_\mu \sigma^\mu_{\alpha\dot{\alpha}} , \qquad \quad
   k^2\!= \det\{k_{\alpha\dot{\alpha}}\} = 0 ,
\label{SpinorsMassless}
\ee
where $\sigma^\mu=(1,\bm{\sigma})$. Assuming
$k^\mu=\omega\;\!(1,\cos\varphi \sin\theta,\sin\varphi \sin\theta,\cos\theta)$,
we may pick
\be
   \ket{k}_\alpha = \sqrt{2\omega}
      \begin{pmatrix}\!-e^{-i\varphi}\sin\frac{\theta}{2} \\
      \cos\frac{\theta}{2} \end{pmatrix}\!, \qquad \quad
   |k]^{\dot{\alpha}} = \sqrt{2\omega}
      \begin{pmatrix} \cos\frac{\theta}{2} \\
      e^{i\varphi}\sin\frac{\theta}{2} \end{pmatrix}\!,
\label{MasslessSpinorSolutionAngles}
\ee
with the understanding that little-group ${\rm U}(1)$ transformations,
$\ket{k} \to e^{-i\phi(k)/2} \ket{k}$,
$|k] \to e^{i\phi(k)/2} |k]$,
leave \eqn{SpinorsMassless} invariant and are thus allowed.

Massive spinors~\cite{Arkani-Hamed:2017jhn}\footnote{For earlier iterations
of the massive spinor-helicity formalism see
\rcites{Kleiss:1986qc,Dittmaier:1998nn,Schwinn:2005pi,
Conde:2016vxs,Conde:2016izb}.}
carry additional ${\rm SU}(2)$ little-group indices~$a$ and obey
\be
   \ket{p^a}_{\;\!\!\alpha}\;\![p_a|_{\dot{\alpha}}
   := \epsilon_{ab} \ket{p^a}_{\;\!\!\alpha}\;\![p^b|_{\dot{\alpha}}
    = p_{\alpha\dot{\alpha}}
   := p_\mu \sigma^\mu_{\alpha\dot{\alpha}} , \qquad \quad
   p^2\! = \det\{p_{\alpha\dot{\alpha}}\} = M^2 \neq 0 .
\label{SpinorsMassive}
\ee
For $p^\mu=(\varepsilon,\rho\cos\varphi \sin\theta,\rho\sin\varphi \sin\theta,\rho\cos\theta)$,
such that $\varepsilon^2-\rho^2=M^2$, one may choose the massive spinors explicitly as (columns are labeled by $a$, rows by $\alpha$ or $\dot{\alpha}$)
\small
\be
\ket{p^a}_\alpha\!=\!
\begin{pmatrix}
\sqrt{\varepsilon\!-\!\rho}\:\!\cos\frac{\theta}{2} &
\!\!-\sqrt{\varepsilon\!+\!\rho}\,e^{-i\varphi}\:\!\!\sin\frac{\theta}{2} \\
\!\sqrt{\varepsilon\!-\!\rho}\,e^{i\varphi}\:\!\!\sin\frac{\theta}{2} &
\sqrt{\varepsilon\!+\!\rho}\:\!\cos\frac{\theta}{2}
\end{pmatrix}\!, ~~
[p^a|_{\dot{\alpha}}\!=\!
\begin{pmatrix}
\!-\sqrt{\varepsilon\!+\!\rho}\,e^{i\varphi}\:\!\!\sin\frac{\theta}{2} &
\!-\sqrt{\varepsilon\!-\!\rho}\:\!\cos\frac{\theta}{2} \\
\sqrt{\varepsilon\!+\!\rho}\:\!\cos\frac{\theta}{2} &
\!\!-\sqrt{\varepsilon\!-\!\rho}\,e^{-i\varphi}\:\!\!\sin\frac{\theta}{2}
\end{pmatrix}\!,
\label{MassiveSpinorSolutionAngles}
\ee
\normalsize
This is ambiguous for $\bm{p}=0$, so one may choose \eg
\be
p^\mu=(M,\bm{0}) = 0 \qquad \Rightarrow \qquad
\bra{p^a}^\alpha = \sqrt{M} \epsilon^{\alpha a} , \qquad
[p^a|_{\dot{\alpha}} = \sqrt{M} \epsilon_{\dot{\alpha} a} .
\label{MassiveSpinorSolutionRest}
\ee
The ${\rm SU}(2)$ little-group rotations,
$\ket{p^a} \to U^a{}_b(p) \ket{p^a}$,
$|p^a] \to U^a{}_b(p) |p^b]$,
leave momentum $p^\mu$ invariant
and correspond to choosing different spin quantization axes~$n^\mu$.
(More details may be found in
\cite{Arkani-Hamed:2017jhn,Ochirov:2018uyq,Aoude:2021oqj}).
The parametrization~\eqref{MassiveSpinorSolutionAngles} picks
$n^\mu = (\rho,\varepsilon\cos\varphi \sin\theta,\varepsilon\sin\varphi \sin\theta,\varepsilon\cos\theta)/M$,
\ie quantization along the momentum,
while \eqn{MassiveSpinorSolutionRest} chooses the conventional $z$-axis.

The momentum spinors serve as basic building blocks for scattering amplitudes.
For massless particles, the spin is always quantized along the momentum,
and is thus counted by helicity weights:
$-1/2$ for each $\ket{k}$ and $+1/2$ for $|k]$.
Moreover, each massive spin-$s$ particle is represented
by $2s$ symmetrized ${\rm SU}(2)$ indices.
We denote the corresponding symmetrized tensor product of spinors
by $\odot$, following \cite{Guevara:2017csg}.

\paragraph{Spherical harmonics in 4d.}
Returning to the spherical harmonics,
we may now embed the 3d construction in 4d.
Namely, we regard it as corresponding
to the default choice of the time direction $u^\mu=(1,\bm{0})$
and the celestial sphere swept by a massless momentum
$k^\mu=\omega\;\!(1,\hat{\bm{k}}(\theta,\varphi))$
and parametrized by the spinors
$\ket{k}_\alpha = \sqrt{2\omega} \kappa_-^{a=\alpha}$ and
$|k]^{\dot{\alpha}} = \sqrt{2\omega} \kappa_+^{a=\dot{\alpha}} $.
Lorentz boosts change the time direction
and induce M\"obius transformations on the celestial sphere.

For a general time direction $u^\mu$ (such that $u^2 = 1$ and $u^0>0$),
we choose to parametrize the celestial sphere by the massless spinors
$\ket{k}_\alpha$ and $|k]^{\dot{\alpha}}$.
Of course, the quantum numbers of a spherical harmonic
must be the same as in the rest frame of $u^\mu$.
The massive spinors $\bra{u^a}^\alpha$ and $[u^a|_{\dot{\alpha}}$
provide a perfect transformation device
between the current inertial frame and the rest frame of $u^\mu$.
This brings us to \eqn{SphHarmSpinors4d}, \ie
\be
{}_h\tilde{Y}_{j,m}(k;u,n) :=
  \frac{1}{\bra{k}u|k]^j}\,
   \lefteqn{ \overbrace{\phantom{[u_{(1} k] \cdots [u_1 k]} }^{j-m} }
   \lefteqn{ \underbrace{\phantom{[u_{(1} k] \cdots [u_1 k]
                                  [u_2 k] \cdots [u_2 k]}\,}_{j+h} }
   [u_{(1} k] \cdots [u_1 k]
   \overbrace{ [u_2 k] \cdots [u_2 k]
               \underbrace{ \braket{k u_2} \cdots \braket{k u_{2)}}}_{j-h}
               }^{j+m} .
\label{SphHarmSpinors4dAlt}
\ee
Here the subscripts $1$ and $2$ are the explicitly symmetrized
little-group indices, and the prefactor involving
$\bra{k}u|k] = 2 k \cdot u \xrightarrow[\bm{u} \to 0]{} 2k^0$
serves to cancel out the mass dimension.
Together with \eqn{MassiveSpinorSolutionRest},
it guarantees
the consistency with the rest-frame definition~\eqref{SphHarmSpinors3d} ---
up to the functional ${\rm U}(1)$
transformation of the form~\eqref{SphHarmRotationU1}
in view of the differences in the $\varphi$-dependence
between \eqns{Spinors3d}{MasslessSpinorSolutionAngles}.
This is an example of acceptable convention discrepancies,
which maybe caused by switching between different spinor parametrizations.
The validity of the harmonics~\eqref{SphHarmSpinors4dAlt}
as representations of the spin algebra
follows from the properties of massive spinors,
see \eg \cite{Guevara:2019fsj,Aoude:2021oqj}.
Note that the dependence on the spin-quantization axis~$n^\mu$
enters via the choice of the massive spinors,
as discussed around \eqn{SpinQuantAxis}.
In other words, the ${\rm SU}(2)$ little-group transformations
$\ket{u^a} \to U^a{}_b(p) \ket{u^a}$, $|u^a] \to U^a{}_b(p) |u^b]$
induce the ${\rm SO}(3)$ rotations of $n^\mu$
orthogonally to the time direction given by~$u^\mu$.
Since the choice of spinors for $u^\mu$ defines $n^\mu$,
the notation may as well be compressed to ${}_hY_{j,m}(k;u)$.

Let us now discuss the orthonormality property~\eqref{SphHarmCovNorm}.
It is valid for the normalized versions of the covariant harmonics,
rescaled from those in \eqn{SphHarmSpinors4dAlt}
analogously to their non-covariant counterparts in \eqn{SphHarmGoldberg}.
It can be easily seen that in the rest frame of~$u^\mu$
the covariant integration measure reduces to the solid-angle one:
\be
\frac{2}{\omega}\!\int\!d^4 k\;\!\delta^+(k^2) \delta(k \cdot u - \omega)
   ~\xrightarrow[\bm{u} \to 0]{}~ \int\!d\Omega_{\hat{\bm{k}}} , \qquad
   k^0 = |\bm{k}| = \omega .
\label{S2MeasureCov}
\ee
So \eqn{SphHarmCovNorm} clearly holds for $\bm{u}=0$,
and what we need is to extend it to any $u^\mu$.

\paragraph{Spinor integration.}
To expose the properties of the measure~\eqref{S2MeasureCov} in a neat way,
we first rewrite it using a null basis \cite{Newman:1961qr}:
\be
k^\mu\!= t \Big( r^\mu + \gamma q^\mu
                 + \frac{z}{2} [r|\bar{\sigma}^\mu\ket{q}
                 + \frac{\bar{z}}{2} [q|\bar{\sigma}^\mu\ket{r} \Big)
~~\Rightarrow~~
\int\!d^4k = \frac{i(r+q)^4\!}{4}\!\int\!t^3
   dt \wedge d\gamma \wedge dz \wedge d\bar{z} ,
\label{NPexpansion}
\ee
where $\bar{\sigma}^\mu=(1,-\bm{\sigma})$,
and the massless vectors $r^\mu$ and $q^\mu$ are not collinear
but otherwise arbitrary.
Adding the masslessness condition eliminates $\gamma$ from the measure:
\be\!\!
\int\!d^4k\;\!\delta^+(k^2) = \frac{i(r+q)^2\!}{4}\!\int_0^\infty\!\!t dt
   \int\!dz \wedge d\bar{z} , \qquad
k^\mu\!= \frac{t}{2} \big(\bra{r} + z \bra{q}\big) \sigma^\mu
   \big(|r] + \bar{z} |q]\big) .
\ee
(Here for concreteness one may assume $r^0, q^0 > 0$ so that $k^0 > 0$.)
However, this massless measure may now be rewritten using spinor integration
\cite{Britto:2005ha,Anastasiou:2006jv,Mastrolia:2009dr}
\be
\int\!d^4 k\;\!\delta^+(k^2) = -\frac{i}{4} \int_0^\infty\!\!t dt
   \int_{\lb = \bar{\la}}\!\braket{\la d\la} \wedge [\lb d\lb] , \qquad
k^\mu\!= \frac{t}{2} \bra{\la}\sigma^\mu|\lb] ,
\ee
such that
the dependence on $r^\mu$ and $q^\mu$ has entirely canceled out due to
\be
(r + q)^2 dz \wedge d\bar{z}
 = -\big(\bra{r} + z \bra{q}\big) \ket{q} dz \wedge
   \big([r| + \bar{z} [q|\big)|q]d\bar{z}
 = -\braket{\la d\la} \wedge [\lb d\lb] .
\ee
Now introducing the second delta function let us fix
the energy scale of $k^\mu$ and get
\be
\frac{1}{\omega}\!\int\!d^4 k\;\!\delta^+(k^2) \delta(k\!\cdot\!u - \omega)
= -i\!\int_{\lb = \bar{\la}}\!\!
   \frac{\braket{\la d\la} \wedge [\lb d\lb]}{\bra{\la}u|\lb]^2} , \qquad
k^\mu\!= \omega \frac{\bra{\la}\sigma^\mu|\lb]}{\bra{\la}u|\lb]} .
\label{S2MeasureCov2Spinor}
\ee

This measure allows us to reformulate
the orthonormality property~\eqref{SphHarmCovNorm}
of the spin-weighted spherical harmonics in the following way:
\be
\int_{\lb = \bar{\la}}\!\!
   \frac{\braket{\la d\la} \wedge [\lb d\lb]}{\bra{\la}u|\lb]^2}\;
   {}_hY^*_{j',m'}(\la,\lb;u)\;{}_hY_{j,m}(\la,\lb;u)
 = \frac{i}{2} \delta_j^{j'} \delta_m^{m'} ,
\label{SphHarmCovNormSpinors}
\ee
where the notation ${}_hY_{j,m}(\la,\lb;u) := {}_hY_{j,m}(k;u)$
serves to emphasize their independence of the energy scale.
Then the validity of \eqn{SphHarmCovNorm} for $\bm{u} \neq 0$ follows from
the fact that the entire left-hand side is independent of $\omega = k \cdot u$.
Indeed, for any spinor conventions and in any frame,
we can rewrite it as the same integral over the complex plane
by parametrizing $\ket{\la} = \ket{u^1} + z \ket{u^2}$
and $|\lb] = |u_1] + \bar{z} |u_2]$,
so that the left-hand side of \eqn{SphHarmCovNormSpinors}
will exclusively involve the following ingredients:
\be
\begin{aligned}
\braket{u_a \la} & = -\delta_a^1\!- \delta_a^2 z , \\
[u_a \lb] & = \epsilon_{1a} + \epsilon_{2a} \bar{z} ,
\end{aligned} \qquad \quad
\begin{aligned}
\braket{\la d\la} \wedge [\lb d\lb] & = -dz \wedge d\bar{z}
 := 2i\,d\Re{z} \wedge d\Im{z} , \\
\bra{\la}u|\lb] & = 1 + z \bar{z} .
\end{aligned}
\ee
Therefore, it only depends on the quantum numbers $h,j,j',m$ and $m'$,
and may only produce a combinatorial result,
which may as well be fixed at $u^\mu = (1,\bm{0})$.

\section{Frame transformations of harmonics}
\label{app:Transformations}

Here we derive the spinor transformations~\eqref{eq:SpinorTransform},
which induce the relationship between
covariant spin-weighted spherical harmonics
${}_h\tilde{Y}_{j,m}(k;u)$ and ${}_h\tilde{Y}_{j,m}(k;v)$.

These harmonics correspond to two different
unit timelike vectors $u^\mu$ and $v^\mu$, with a relative Lorentz factor
\be
\gamma := u \cdot v =: \frac{1}{\sqrt{1-\nu^2}} , \qquad \quad
0 \leq \nu < 1 .
\ee
These vectors can be Lorentz-transformed into each other
using the minimal boost
\be
L^\rho{}_\sigma(v\!\leftarrow\!u)
:= \delta^\rho_\sigma + 2v^\rho u_\sigma
 - \frac{(u+v)^\rho (u+v)_\sigma\!}{1 + u \cdot v}
 = \exp\!\Big( \tfrac{i \log(\gamma + \sqrt{\gamma^2-1})}{\sqrt{\gamma^2-1}}
   u^\mu v^\nu \Sigma_{\mu\nu}\!\Big)^\rho_{~\sigma} ,
\label{eq:MinBoost}
\ee
written in terms of the spin-1 Lorentz generators
$(\Sigma^{\mu\nu})^\rho{}_\sigma
:= i [\eta^{\mu\rho} \delta^\nu_\sigma - \eta^{\nu\rho} \delta^\mu_\sigma]$.
The spinors may be boosted
using the corresponding ${\rm SL}(2,\mathbb{C})$ transformations, namely
\be
S^\alpha{}_\beta(v\!\leftarrow\!u)
 = \exp\!\big( \tfrac{i \log\mu}{\gamma \nu} u^\mu v^\nu \sigma_{\mu\nu}
         \big)^\alpha{}_\beta , \qquad \quad
\mu := \gamma + \sqrt{\gamma^2-1} ,
\ee
written in terms of the chiral spin-1/2 generators
$\sigma^{\mu\nu} := \frac{i}{2} \sigma^{[\mu} \bar{\sigma}^{\nu]}$.
Using the Clifford-algebra property
$\sigma^{(\mu} \bar{\sigma}^{\nu)} = \eta^{\mu\nu}$,
it is easy to derive
\beal
\big( \tfrac{i \log\mu}{\gamma \nu} u^\mu v^\nu \sigma_{\mu\nu}
\big)^{2n} \ket{u^a} & = (\log\!\sqrt{\mu})^{2n} \ket{u^a} , \\
\big( \tfrac{i \log\mu}{\gamma \nu} u^\mu v^\nu \sigma_{\mu\nu}
\big)^{2n+1} \ket{u^a} & = ({-}\log\!\sqrt{\mu})^{2n+1}
   \Big( \tfrac{1}{\nu} \ket{u^a} - \tfrac{1}{\gamma\nu} |v|u^a] \Big) . 
\eeal
This lets us sum the matrix exponent, whose action simplifies to
\beal
S^\alpha{}_\beta(v\!\leftarrow\!u) \ket{u^a} &
 = \frac{\sqrt{\mu}}{\mu+1} \Big( \ket{u^a} + |v|u^a] \Big) .
\eeal
We thus arrive at the following massive-spinor transformations:
\be
\ket{v^b} = \frac{\sqrt{\mu}}{\mu+1}
   U^b{}_a(v\!\leftarrow\!u) |u\!+\!v|u^a] , \qquad \quad
|v^b] = \frac{\sqrt{\mu}}{\mu+1}
   U^b{}_a(v\!\leftarrow\!u) |u\!+\!v\ket{u^a} .
\label{eq:SpinorTransformFull}
\ee
Here we have allowed for the ${\rm SU}(2)$ matrix $U^b{}_a(v\!\leftarrow\!u)$.
Its purpose is to fix the misalignment between
what we get from the minimal boost~\eqref{eq:MinBoost}
and the desired spin quantization axis for the resulting time direction,
which generically do not coincide:
\be
n_v^\mu := \frac{1}{2}
   (\bra{v_2}\sigma^\mu|v^2] + [v_2|\bar{\sigma}^\mu\ket{v^2})
~~\neq~~ L^\mu{}_\nu(v\!\leftarrow\!u) n^\nu
 = n^\mu - \frac{n \cdot v}{1+u\cdot v} (u+v)^\mu .
\label{eq:SpinMisalignment}
\ee
In fact, unitary matrices like $U^b{}_a(v\!\leftarrow\!u)$ represent
the ${\rm SO}(3)$ rotations of the spin quantization axis
even in the absence of Lorentz-frame boosts.
Therefore, the spinor transformations~\eqref{eq:SpinorTransformFull}
induce the most general frame transformations
of the covariant spherical harmonics.

\bibliographystyle{JHEP}
\bibliography{references}
\end{document}